\documentclass{jpp}
\usepackage{graphicx}
\usepackage{epstopdf,epsfig}
\usepackage{hyperref}
\usepackage[normalem]{ulem}

\usepackage{xcolor}

\def \bm {\mathbf}

\shorttitle{On the distortion of the ion velocity distribution functions}
\shortauthor{S. Perri et al.}

\title{On the deviation from Maxwellian of the ion velocity distribution functions in the turbulent magnetosheath}

\author{S. Perri\aff{1}
  \corresp{\email{silvia.perri@fis.unical.it}},
  D. Perrone\aff{2,3},
  E. Yordanova\aff{4},
  L. Sorriso-Valvo\aff{5,6},
  W. R. Paterson\aff{7},
  D. J. Gershman \aff{7},
  B. L. Giles \aff{7},
  C. J. Pollock \aff{7},
  J. C. Dorelli \aff{7},
  L. A. Avanov \aff{7},
  B. Lavraud \aff{8},
  Y. Saito \aff{9},
  R. Nakamura \aff{10},
  D. Fischer \aff{10},
  W. Baumjohann \aff{10},
  F. Plaschke \aff{10},
  Y. Narita \aff{10},
  W. Magnes \aff{10},
  C. T. Russell \aff{11},
  R. J. Strangeway \aff{11},
  O. Le Contel \aff{12},
  Y. Khotyaintsev \aff{4},
 \and F. Valentini\aff{1}}

\affiliation{\aff{1}Dipartimento di Fisica, Universit\`a della Calabria, Arcavacata di Rende, Italy
\aff{2}ASI - Italian Space Agency, Rome, Italy
\aff{3}Department of Physics, Imperial College London, London SW7 2AZ, UK
\aff{4}Swedish Institute of Space Physics, Uppsala, Sweden
\aff{5}Departamento de F\'isica, Escuela Polit\'ecnica Nacional, Quito, Ecuador
\aff{6}Nanotec/CNR, U.O.S. di Rende, Arcavacata di Rende, Italy
\aff{7} NASA Goddard Space Flight Center, Greenbelt, USA
\aff{8} Institut de Recherche en Astrophysique et Plan\'etologie, CNRS, UPS, CNES, Universit\'e de Toulouse, Toulouse, France
\aff{9} JAXA, Tokyo, Japan
\aff{10} Space Research Institute, Austrian Academy of Science, Graz, Austria
\aff{11} Institute of geophysics and planetary physics, University of California, Los Angeles, USA
\aff{12}Laboratoire de Physique des Plasmas CNRS/Ecole Polytechnique/Sorbonne Universit\'e/Universit\'e Paris-Sud/Observatoire de Paris, Paris, France
}
\begin{document}

\maketitle

\begin{abstract}
The degree of deviation from the thermodynamic equilibrium in the ion velocity distribution functions (VDFs), measured by the Magnetospheric Multiscale (MMS) mission in the Earth's turbulent magnetosheath, is quantitatively investigated. Taking advantage of MMS ion data, having a resolution never reached before in space missions, and of the comparison with Vlasov-Maxwell simulations, this analysis aims at relating any deviation from Maxwellian equilibrium to typical plasma parameters. 
Correlations of the non-Maxwellian features with plasma quantities such as electric fields, ion temperature, current density and ion vorticity are very similar in both magnetosheath data and numerical experiments, and suggest that distortions in the ion VDFs occur close to (but not exactly at) peaks in current density and ion temperature. Similar results have also been found during a magnetopause crossing by MMS. This work could help clarifying the origin of distortion of the ion VDFs in space plasmas.
\end{abstract}

\section{Introduction}\label{sec_intro}
The interplanetary space is permeated by a plasma in which the effects of particle collisions can be considered negligible. In such a nearly collisionless medium, in the range of scales where kinetic effects dominate the plasma dynamics, the particle velocity distribution functions (VDFs) are observed to be far from the thermodynamic equilibrium.
Moreover, recent self-consistent numerical simulations and nonlinear models of plasma turbulence dynamics have shown the emergence of significant non-Maxwellian features in the particle VDFs. In particular, 2D-3V (two dimensions in physical space and three dimensions in velocity space) kinetic, hybrid Vlasov-Maxwell (HVM) numerical simulations \citep{Valentini07}, which reproduce the turbulent energy cascade down to ion scales, have highlighted significant local departures of the ion VDFs from a Maxwellian shape close to current sheets, generated by the turbulent cascade and non-uniformly distributed through space, which are often associated with the observed ion temperature anisotropy \citep{Servidio12,Greco12}. A certain degree of correlation between the emergence of non-Maxwellian features in the ion VDFs and the presence of regions of high ion vorticity has also been found in such simulations \citep{Valentini16,Sorriso2018}. Localized heating close to strong enhancements of ion vorticity has also been observed in other numerical experiments \citep{Franci2016,Parashar2016}. Finally, distortions of the VDFs have been found to be related to strong magnetic field activity and to the magnetic field topology (see \citet{Servidio12}). 
On the other hand, recent analysis made on solar wind and near-Earth plasma have shown that at proton scales the plasma is characterized by magnetic discontinuities \citep{Retino07,Perri12} that are connected through spatial scales from ion to electron scales \citep{Greco16,Perrone2016,Perrone2017}, giving rise to a complex network, whose effect on the plasma dynamics is still under investigation. Local heating and magnetic energy dissipation at such thin current sheets have been observed in numerical simulations \citep{Wu2013,Perrone2014,Wan15,Sorriso2018b}. Moreover, very recent works have also found that velocity space fluctuations of the ion VDF are characterized by a broad, power-law Hermite spectrum (that is an enstrophy cascade in the velocity space) both in {\it in situ} data \citep{Servidio17} and in numerical simulations \citep{Pezzi2018} and that velocity-space activity is intermittent and concentrated near coherent structures \citep{Pezzi2018}.

Thanks to the launch of the Magnetospheric Multiscale (MMS) mission we get both the advantage of very high resolution plasma measurements and the possibility of spatial measurements from a constellation of four identical satellites able to approach a minimum distance close to $7$ km. Thus, a big step forward to the understanding of the kinetic physics at both ion and electron discontinuities has been made. Indeed, observations of reconnecting current sheets in the Earth magnetopause have been reported \citep{Burch16, Ergun16}, as well as signatures of crossings of ion and electron diffusion regions in the Earth magnetosheath \citep{Eriksson16, Yordanova16, Voroes17} and in the magnetopause \citep{Lavraud16}. Those observations clearly show the presence of magnetic energy conversion to plasma heating and to signatures of departure of ion and electron VDFs from the local thermodynamic equilibrium \citep{Graham17,Sorriso2019}. Thus, one question can be whether the departure from the Maxwellian equilibrium in a plasma at kinetic scales is statistically significantly related to specific plasma quantities. In order to try to answer such question, here we report a statistical investigation of the correlation between the degree of departure of the ion VDF from a Maxwellian shape and the plasma characteristic parameters, such as temperature anisotropy, current density, and ion vorticity, by comparing MMS data intervals in the Earth's magnetosheath and HVM simulations of turbulence.

\section{Magnetosheath event overview and measure of deviation from Maxwellian plasma}\label{sec_event}

In this work, we make use of the high resolution (150 ms) ion VDFs from the Fast Plasma Investigation (FPI) instrument on board MMS \citep{Pollock16}, the magnetic field from the merged fluxgate (FGM) \citep{Russell16} and the search coil (SCM) data \citep{LeContel16}, at about 1kHz resolution \citep{Fischer16}, and the electric field data from Electric Double Probes (EDP) instrument, at about $8$ kHz sampling rate \citep{Torbert16,Ergun16, Lindqvist16}. The data describe a $5$ minute period on 2015 November 30 (from 00:21 to 00:26 UT) where the MMS spacecraft were immersed in the quasiparallel turbulent magnetosheath (see \citet{Yordanova16,Voroes17}). In this interval, the mean magnetic field (averaged over the whole 5 minutes interval) is $B_0\sim 44$ nT, the ion to electron temperature ratio $T_i/T_e \sim 7$ and the plasma beta (namely the ratio between the plasma kinetic pressure and the magnetic pressure) shows large amplitude fluctuations and is greater than $1$.
The aim of this work is to investigate the possible presence of non-Maxwellian features in the ion VDFs close to intermittent magnetic structures by means of a parameter that quantifies the deviation of the measured VDF from a Maxwellian shape \citep{Greco12}. 
Then, these distortions will be compared with the ones observed in numerical experiments of turbulence to shed light on the 
physical mechanisms that could play a role in this process.

\begin{figure*}
\centering
\includegraphics[width=25pc]{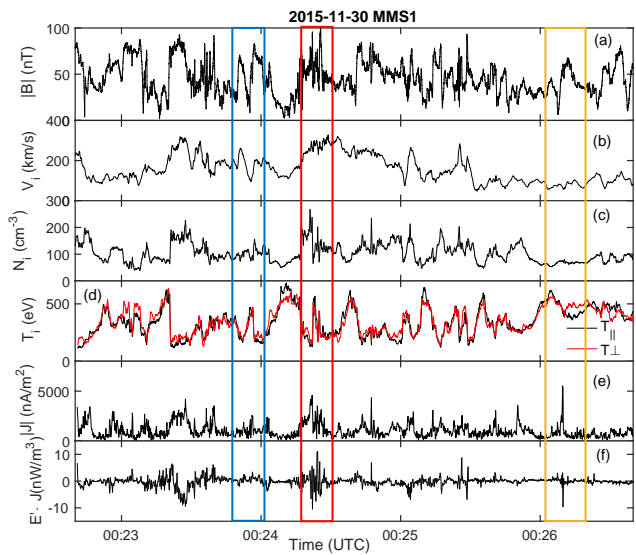}
\caption{From top to bottom: (a) magnetic field magnitude, (b) ion bulk speed, (c) ion density, (d) parallel (black line) and perpendicular (red line) ion temperature, (e) current density magnitude computed using the FPI $150$ ms data, (f) $\mathbf{E}^{'}\cdot \mathbf{J}$, being $\mathbf{E}^{'}$ the electric field in the electron rest frame (i.e., $\mathbf{E}^{'}=\mathbf{E}+(\mathbf{V}_{e}\times \mathbf{B})$) at $150$ ms resolution. The color boxes highlight previously reported studies \citep{Eriksson16,Yordanova16,Voroes17}.}
\label{fig:event}
\end{figure*}

\begin{figure*}
\centering
\includegraphics[width=22pc]{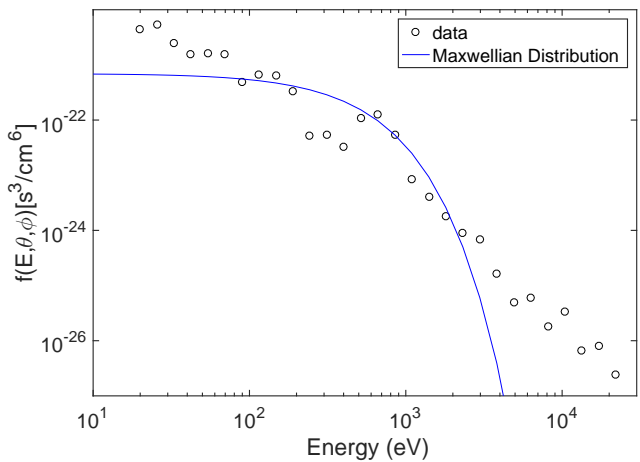}
  \caption{Measured ion velocity distribution function (circles) as a function of energy measured at fix angles and at a given time, compared with the associated Maxwellian distribution (solid line).}
\label{fig:vdf}
\end{figure*}

The event overview is shown in figure \ref{fig:event}. From top to bottom we have (a) the magnetic field magnitude; (b) the ion bulk speed; (c) the ion density; (d) the ion temperature parallel (black line) and perpendicular (red line) to the mean magnetic field $B_0$, (e) the intensity of the current density at $150$ ms resolution, computed as $\mathbf{J}=Nq(\mathbf{V}_i-\mathbf{V}_e)$ with $N$ being the plasma density ($N=N_i \sim N_e$); and (f) the $\mathbf{E'}\cdot \mathbf{J}$, which represents the work done by the fields on particles with $\mathbf{E'}=\mathbf{E}+\mathbf{V}_e\times \mathbf{B}$ in the electron frame. All the data is from the MMS1 spacecraft. This represents a sample of turbulent plasma where three thin current sheets encounters have been analyzed in previous studies \citep{Eriksson16,Yordanova16,Voroes17} and are highlighted with colored boxes in figure \ref{fig:event}. 
Thus, in such a turbulent environment the ion VDFs can be highly distorted and modified as observed in numerical simulations \citep{Servidio12,Perrone13} and in {\it in situ} data \citep{Voroes17,Servidio17,Sorriso2019}.
An example of one-dimensional VDF cut as a function of energy is shown in figure \ref{fig:vdf} (symbols), where a clear departure from a Maxwellian distribution (solid line) is observed. Notice that the deviation is highly significant, since the associated noise level in the VDF in the magnetosheath is very low. 

In order to quantify the statistical occurrence of deviation from Maxwellian, we make use of the parameter \citep{Greco12,Valentini16}
\begin{equation}
\epsilon_i=\frac{1}{n_i}\sqrt{\int (f_i-g_i)^2 d^3v}\, ,
\label{epsilon}
\end{equation}
being $n_i$ the ion density, $f_i$ the observed VDF for the ions, and $g_i$ the equivalent Maxwellian distribution with the same density, temperature, and velocity as the observed one. Thus, $\epsilon_i=0$ indicates Maxwellian VDF, while any $\epsilon_i \ne 0$ suggests a deviation from the equilibrium.  
As pointed out in \citet{Servidio17}, $\epsilon_i^2$ is related to the phase-space enstrophy.
We are comparing the real ion VDFs in the magnetosheath with a Maxwellian distribution as a quantitative measure of distortion of the distributions. We do not actually expect thermodynamic equilibrium in such an environment, so that finding non-Maxwellian distributions is not surprising. For example, the presence of cold ions (few tens of eV) of ionospheric origin \citep{toledo2016,toledo2017,Li2017} could also produce a distortion of the ion VDF from a simple Maxwellian. However, the computation of $\epsilon_i$ allows us to have a quantitative  departure from the Maxwellian equilibrium. 

\begin{figure*}
\centering
\includegraphics[width=22pc]{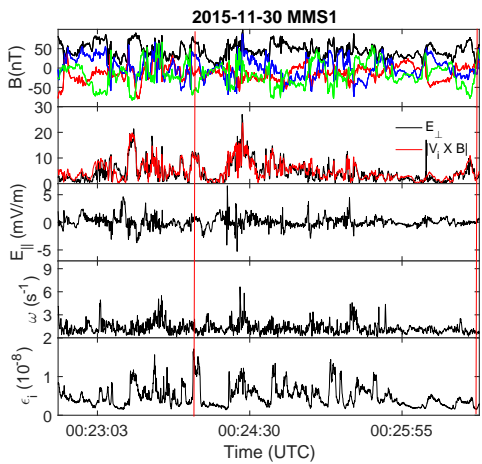}
  \caption{From top to bottom: magnetic field components in the GSE reference frame along with the magnetic field intensity (black line); the electric field perpendicular to the local magnetic field direction (black line) along with the intensity of the $\mathbf{V}_i \times \mathbf{B}$ (red line); the electric field component parallel to the local mean field; the ion vorticity; the derived time series of the $\epsilon_i$ parameter. The two vertical lines indicate (i) a region with $\epsilon_i$ above a $1 \sigma$ threshold that corresponds also to enhancements in the perpendicular electric field and in the ion vorticity, and (ii) a region of very low $\epsilon_i$.}
\label{fig:eps}
\end{figure*}

The time series of $\epsilon_i$ in our turbulent interval is displayed in the bottom panel in figure \ref{fig:eps}; notice the burstiness of this quantity with large deviations sparse throughout the whole interval. 
Besides the time series of $\epsilon_i$, figure \ref{fig:eps} shows, from top to bottom, the magnetic field components in the geocentric-solar-ecliptic (GSE) coordinates ($B_x$-red line; $B_y$-blue line; $B_z$-green line) along with the magnitude of the magnetic field (black line), the electric field $E_{\perp}=\sqrt{E_{\perp1}^2+E_{\perp2}^2}$ perpendicular to the local magnetic field (black line)--where $E_{\perp1}$ lies along $\mathbf{E}\times \mathbf{B}$ and $E_{\perp2}$ is aligned to $\mathbf{B}\times \mathbf{E}\times \mathbf{B}$--together with the $|\mathbf{V}_i \times \mathbf{B}|$ (red line), the electric field component parallel to the local mean field, and the ion vorticity $|\omega|=|\nabla \times \mathbf{V}_i|$ from the 4 spacecraft measurements (the inter-satellite distance in this period is about $10$ km). Notice that ions are magnetized since $E_{\perp}$ and $|\mathbf{V}_i \times \mathbf{B}|$ track each other very well, and the time series of $\epsilon_i$ exhibits features that tend to correlate with $E_{\perp}$. The two red vertical lines in figure \ref{fig:eps} indicate (i) a region where $\epsilon_i$ is above a $1 \sigma$ threshold and seems to be correlated with enhancements in $E_{\perp}$ and $|\omega|$, and (ii) a region (near the right end of the interval) where $\epsilon_i$ is closer to zero (below the $1 \sigma$). 
To understand what causes large deviations from Maxwellian in this plasma sample, we have looked at the 2D VDF cuts in the regions indicated by the vertical lines in figure \ref{fig:eps}. 

\begin{figure*}
\centering
\includegraphics[angle=-90,width=22pc]{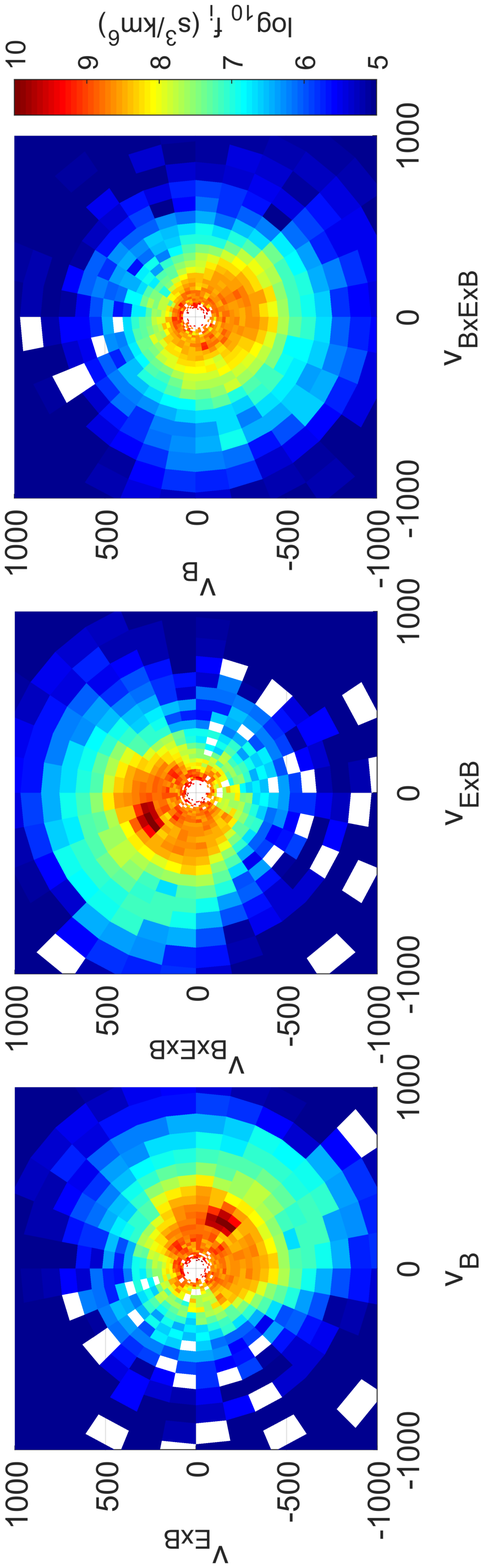}
\includegraphics[angle=-90,width=22pc]{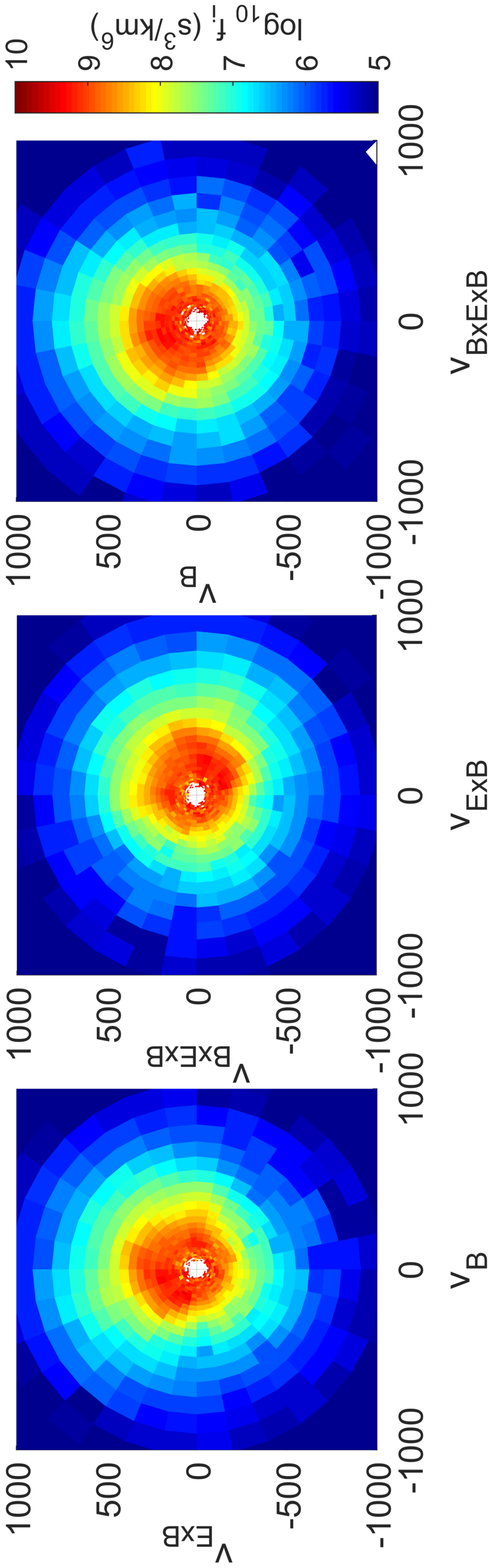}
  \caption{2D VDF cuts in the two regions highlighted with vertical red dashed lines in figure \ref{fig:eps}. The top panels refer to a peak in $\epsilon_i$, while the bottom panels to a valley in $\epsilon_i$. Notice the presence of a particle beam almost aligned with the $\mathbf{B}$ direction in the top panels.}
\label{fig:vdfcut}
\end{figure*}

Figure~\ref{fig:vdfcut} shows the ion VDFs, plotted in a reference frame with one direction aligned to $\mathbf{B}$, one direction parallel to $\mathbf{E}\times \mathbf{B}$, and the third perpendicular to both. It is worth stressing that most of the ion VDFs are characterized by a beam-like distribution (mainly along $\mathbf{B}$), with the beam traveling at about $300$ km/s, which is about $3$ times the local ion Alfv\'en speed \citep{Yordanova16,Sorriso2019}. Such feature has been found to be frequently present in the peaks of $\epsilon_i$ above the threshold (top panels). Figure \ref{fig:vdfcut} also shows the 2D VDFs at a time during which $\epsilon_i$ is below the threshold (bottom panels). Within the interval, the shape changes abruptly (see also \citet{Servidio17}), with the presence also of a more isotropic distributions. In the process of selecting ion VDFs at the peaks in $\epsilon_i$, we noticed that in several cases the ion VDFs are characterized by several ``holes'' in different energy channels and look directions, due to absence of detected particles. Such holes would artificially amplify the difference with a Maxwellian distribution. Therefore we take a special care in our analysis and we compute the $\epsilon_i$ parameter only when ion VDFs values are fully defined.

\begin{figure*}
\centering
\includegraphics[width=10pc]{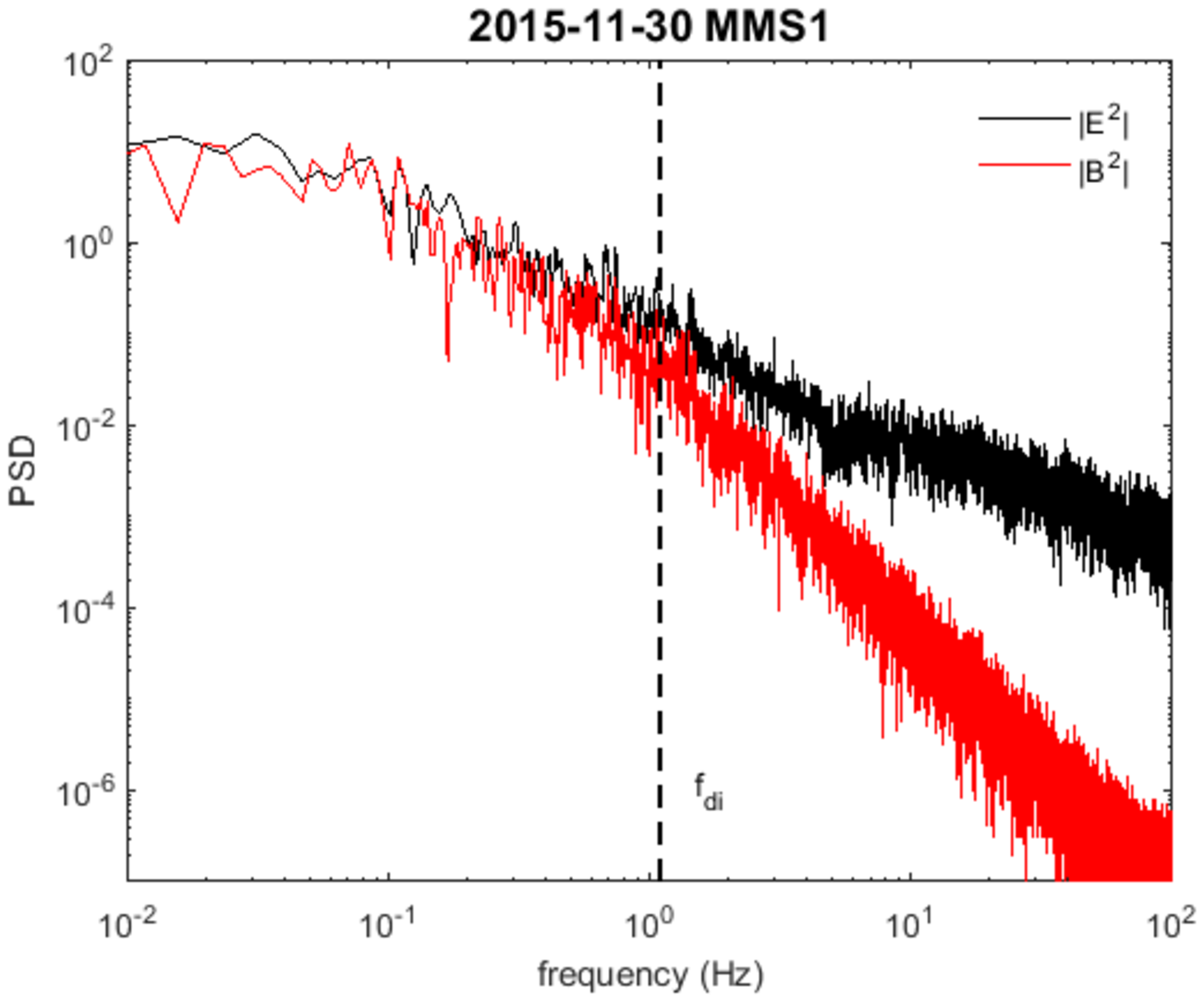}
\includegraphics[width=10pc]{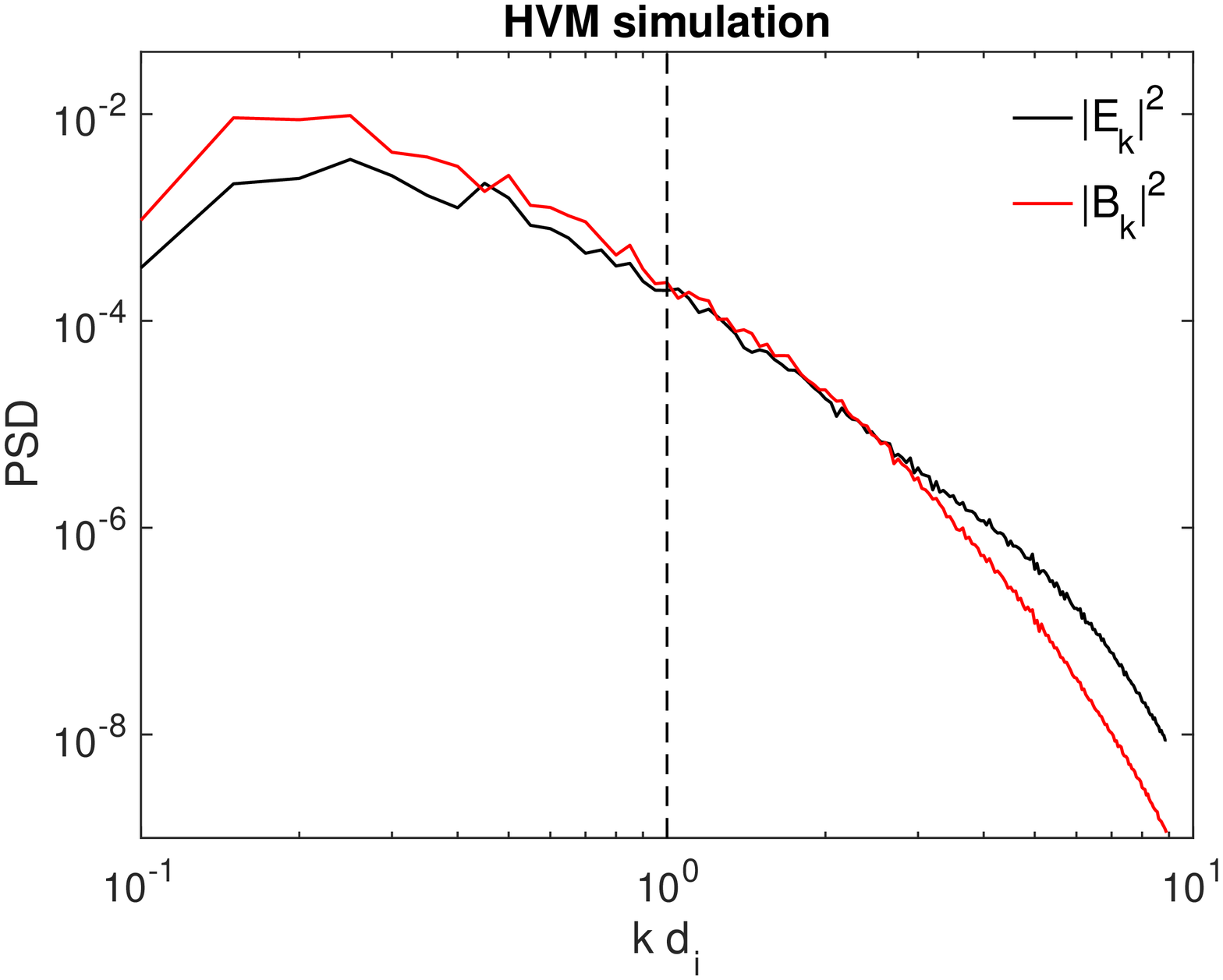}
  \caption{Power spectral densities of the normalized electric (black lines) and magnetic (red lines) fluctuations in the MMS1 data set (left panel) and in the HVM simulation (right panel).}
\label{fig:spectraEB}
\end{figure*}

\section{Comparison between MMS1 data and HVM simulations}\label{corr}
In order to investigate the properties of the regions of high deformation for the ion VDFs, we have compared the results coming from the MMS data with those obtained from a numerical simulation of decaying turbulence with guide field for a collisionless plasma in a 2D-3V phase space domain with the HVM code \citep{Valentini07, Perrone11}. The code solves the Vlasov equation for the ion VDF, while the electrons are considered as an isothermal, massless fluid, and their contribution is taken into account through a generalized Ohm’s law that retains the Hall term and the electron pressure. The Vlasov equation and the Ohm’s law are coupled with the Maxwell equations, where the displacement current is neglected. Quasi-neutrality is assumed. The dimensionless HVM equations can be read as
\begin{eqnarray}
\label{eq:vlas}
\frac{\partial f}{\partial t} + \bm{v} \cdot \frac{\partial f}{\partial \bm{r}}+\left( \bm{E} + \bm{v} \times \bm{B} \right) 
\cdot \frac{\partial f}{\partial \bm{v}} = 0 \\
\label{eq:ohm}
\bm{E} = -\bm{u} \times \bm{B} + \frac{1}{n} \bm{j} \times \bm{B} - \frac{1}{n} \nabla p_e + \eta \bm{j} \\
\label{eq:induction}
\frac{\partial \bm{B}}{\partial t} = - \nabla \times \bm{E}, 
\end{eqnarray}
where $f\equiv f(\bm{r},\bm{v},t)$ is the ion VDF, $\bm{E}(\bm{r},t)$ and $\bm{B}(\bm{r},t)$ are the electric and the magnetic field, respectively, and $\bm{j} = \nabla \times \bm{B}$ is the total current density. The ion density, $n$, and the bulk velocity, $\bm{u}$, are evaluated as velocity moments of the ion VDF. In the above equations, time is scaled by the inverse proton-cyclotron frequency, $\Omega_{ci}^{-1}$, velocity 
by the Alfv\'en speed $v_A = B_0/\sqrt{4\pi n_0 m_i}$, lengths by the ion skin depth, $d_i = v_A/\Omega_{ci}$, and masses by the ion mass, $m_i$.
In this paper, the ion distribution function studied from HVM simulations is actually the proton distribution function, so that hereafter when referring to ions, we mean protons in the simulations.
At t=0, the equilibrium consists of a homogeneous plasma embedded in a uniform background out-of-plane magnetic field, $\mathbf{B}_0$, along the $z$-direction. Ion VDF is initialized with a Maxwellian with homogeneous density. The system evolution is investigated in a double periodic domain $(x,y)$ perpendicular to $\mathbf{B}_0$.
The equilibrium configuration is perturbed by a 2D spectrum of Fourier modes \citep{Servidio12, Perrone13, Valentini14, Valentini16}. The root mean square of the magnetic perturbations is $\delta b/B_0 \sim 0.3$ and neither density disturbances nor parallel fluctuations are imposed at $t=0$. The plasma beta is $\beta = 2v_{th,i}^2/v_A^2 = 0.5$, where $v_{th,i} = \sqrt{T_i/m_i}$ is the ion thermal speed. The ion to electron temperature ratio is $T_i/T_e =1$.  Finally, the system size in the spatial domain is $L = 2\pi \times 20d_i$ in both $x$ and $y$ directions, discretized with $512^2$ grid-points, while the 3D velocity domain, limited by $\pm 5 v_{th,i}$ in each directions, is discretized with an uniform grid of $71^3$ points. 

The omni-directional electric and magnetic field power spectral densities in the simulations are displayed in the right panel of Figure \ref{fig:spectraEB} at the maximum of the turbulent activity. 
At sub-ion scales, the turbulence becomes dominated by the electric fluctuations. The same trend has also been found in the MMS1 data (see left panel in Figure \ref{fig:spectraEB}), where the trace of the  power spectral densities of the normalized electric ($E^2/(v_AB_0)^2$) and magnetic ($(B/B_0)^2$) fluctuations are shown. Indeed, at the frequency corresponding at the ion skin depth in the data interval (vertical dashed line in the left panel in \ref{fig:spectraEB}), the normalized electric field fluctuations have a power higher than the one stored in the magnetic field fluctuations. 

\begin{figure*}
\centering
\includegraphics[width=22pc]{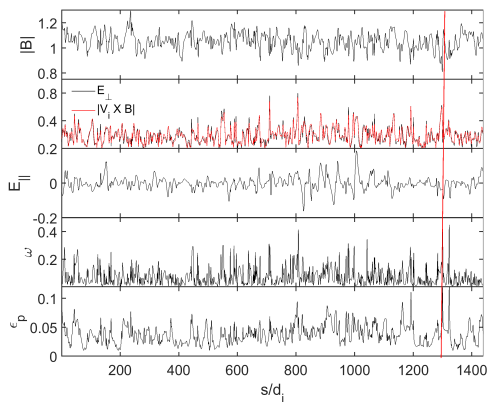}
  \caption{Same format as figure~\ref{fig:eps}. The time series are measured along a one-dimensional cut in the HVM box. The vertical line indicates a region of high vorticity, large value of $\epsilon_p$, and high perpendicular electric field.}
\label{fig:eps_HVM}
\end{figure*}

\begin{figure*}
\centering
\includegraphics[width=10pc]{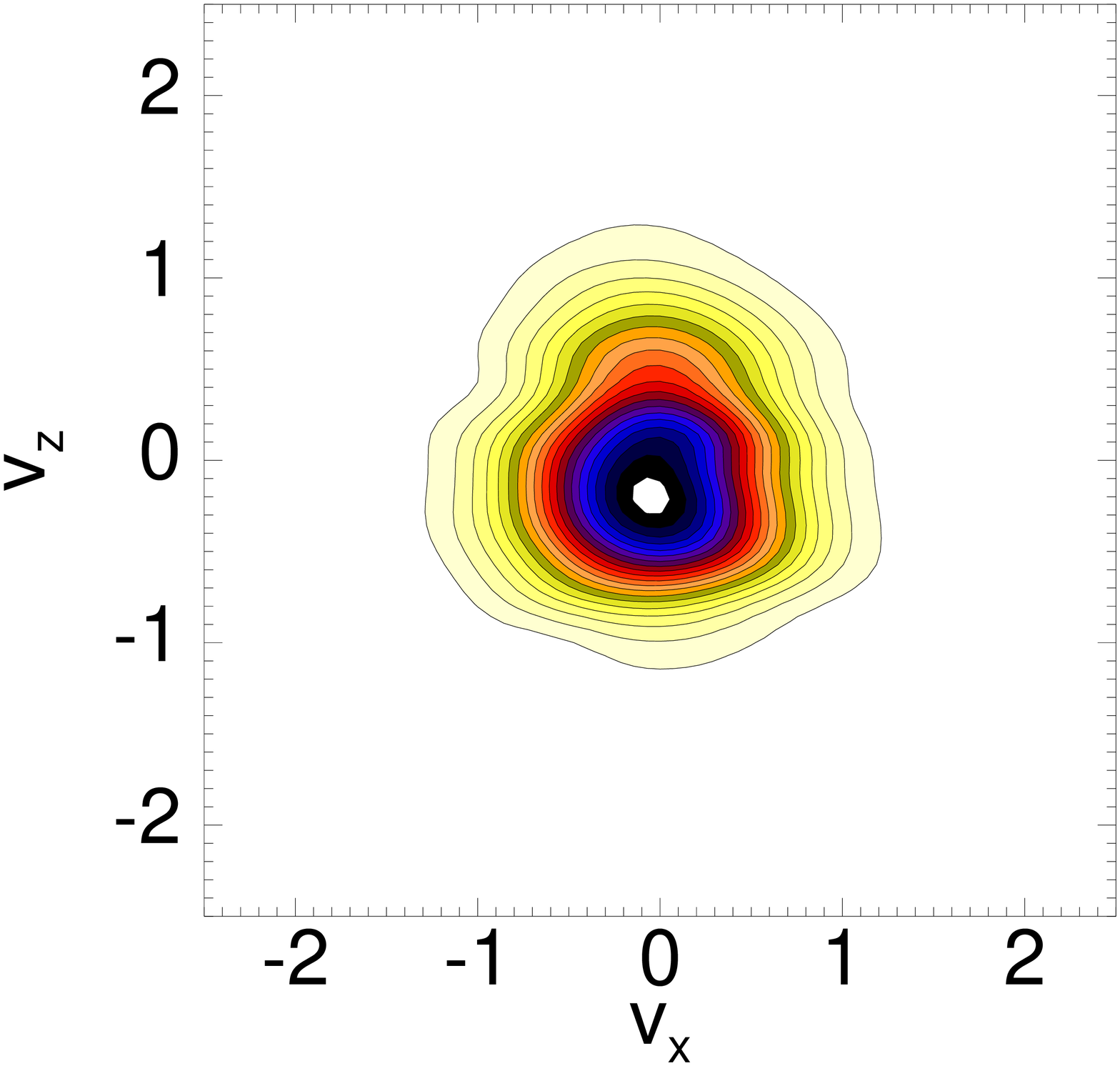}
\includegraphics[width=10pc]{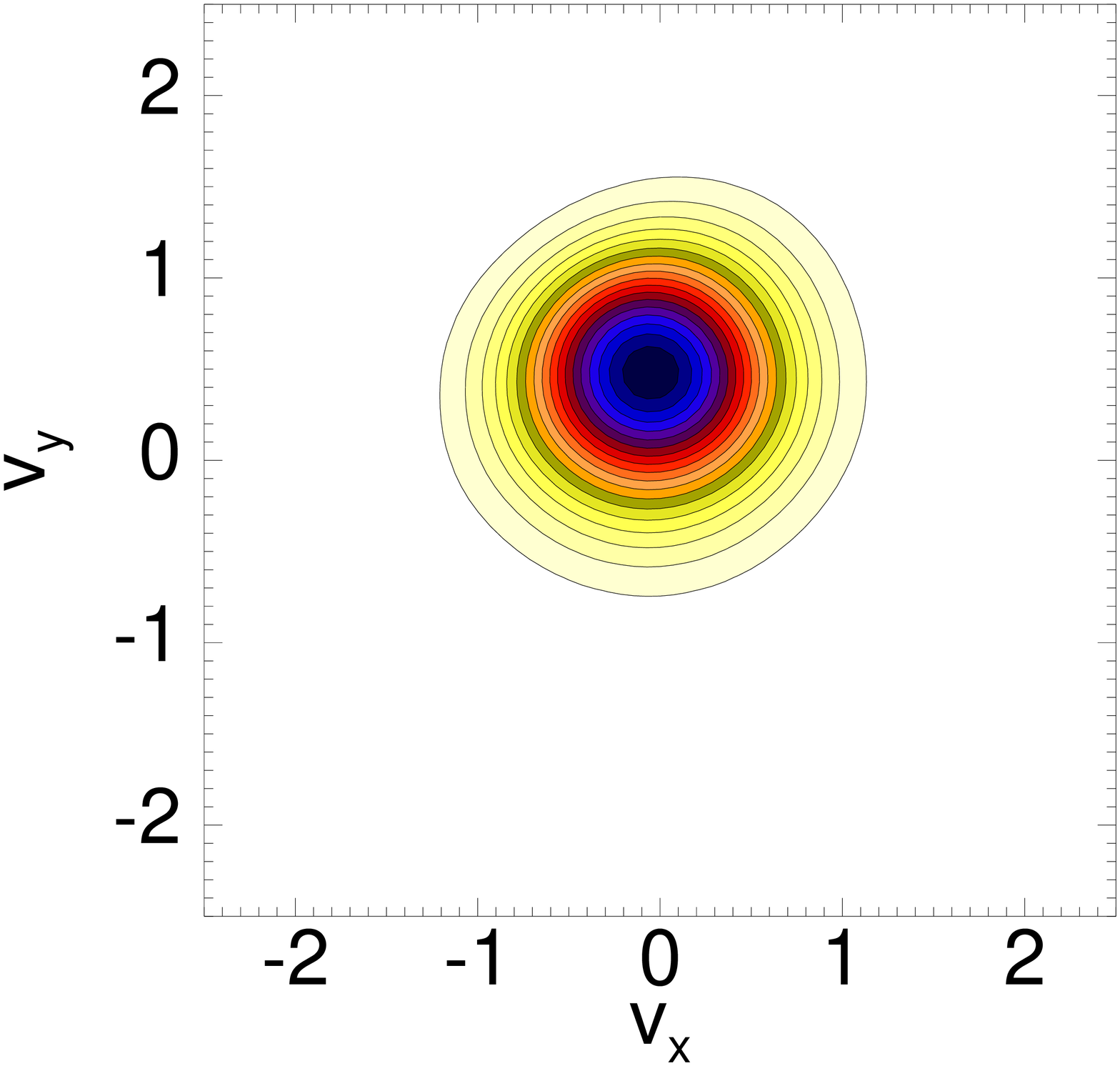}
\includegraphics[width=10pc]{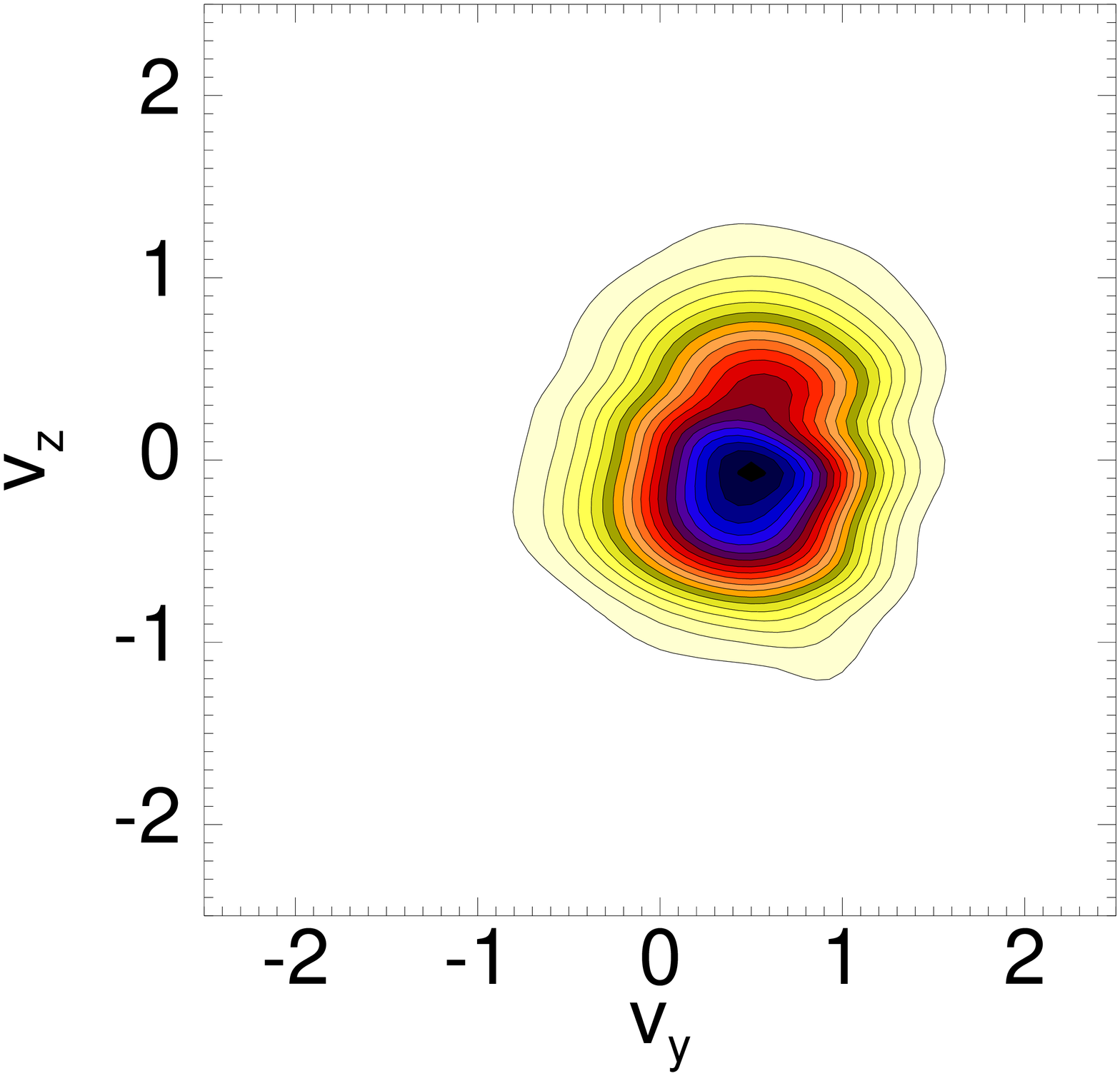}
   \caption{2D VDF cuts in the portion of the signal highlighted in figure \ref{fig:eps_HVM} by the vertical dashed line. A beam almost aligned with the $\mathbf{B}_0$ direction is clearly visible.}
\label{fig:vdfcut_HVM}
\end{figure*}

Figure \ref{fig:eps_HVM} shows an overview of the numerical results, in the same format as figure \ref{fig:eps}. These physical quantities have been tracked along a one-dimensional cut of the 2-D spatial domain of the simulation, i.e. a diagonal path $s$, normalized to $d_i$, that crosses the simulation box several times.
 In the simulation, like in the MMS data, protons are magnetized and $\epsilon_p$ is burst-like. To make a comparison with the VDFs observed by MMS1, we have selected a small portion of the signal where peaks in $\epsilon_p$, $\omega$, and $E_{\perp}$ occur almost simultaneously (i.e., at $s/d_i=1324$, indicated by the vertical line in figure \ref{fig:eps_HVM}) and we have checked the shape of the ion VDF. In figure \ref{fig:vdfcut_HVM}, we plot the 2D cuts of the ion VDF, where the $v_z$ component is parallel to the background magnetic field direction. While in the plane perpendicular to $\mathbf{B}_0$ the VDF is almost isotropic, with a shift towards positive values of $v_y$ due to large scale fluctuations, in the other two planes a beam nearly aligned with $\mathbf{B}_0$ is evident, a feature similar to the one observed in the top panels of figure \ref{fig:vdfcut}. Thus, the presence of high vorticity regions, with enhanced perpendicular electric fields, exhibit distorted ion VDFs with the formation of a beam travelling along $\mathbf{B}_0$ at almost the Alfv\'en speed \citep{Sorriso2019}.

\begin{figure*}
\centering
\includegraphics[width=15pc]{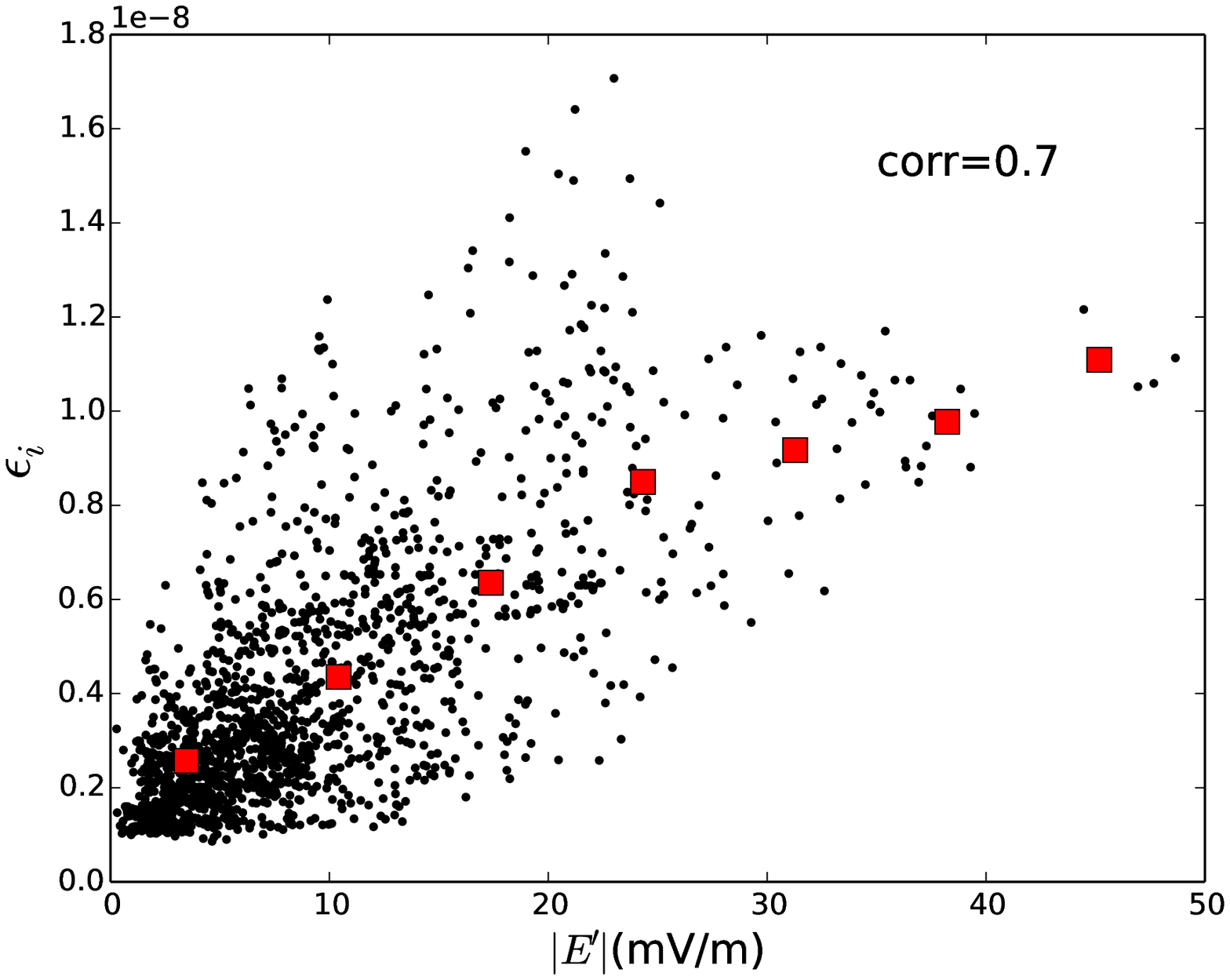}
\includegraphics[width=15pc]{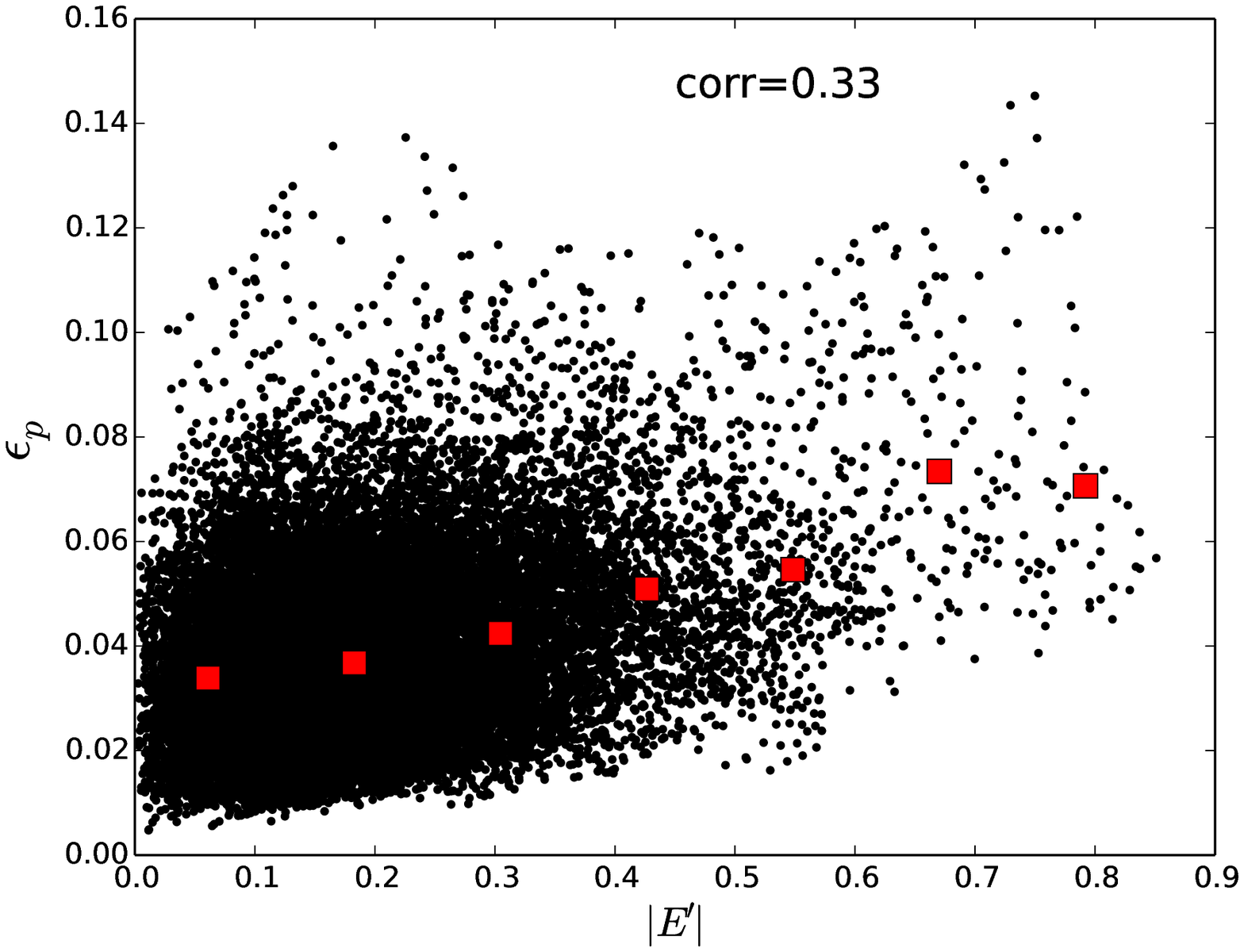}
  \caption{The $\epsilon_i$ parameter as a function of the magnitude of the electric field in the plasma frame in the MMS data (left panel), and in the HVM simulation (right panel). The electric field data in the magnetosheath have been averaged out to $0.15$ sec resolution. The Pearson correlation indices are indicated in the panels. Red squares report the mean values of $\epsilon_i$ within bins of $|E'|$.}
\label{fig:eps-fields}
\end{figure*}

\begin{figure*}
\centering
\includegraphics[width=15pc]{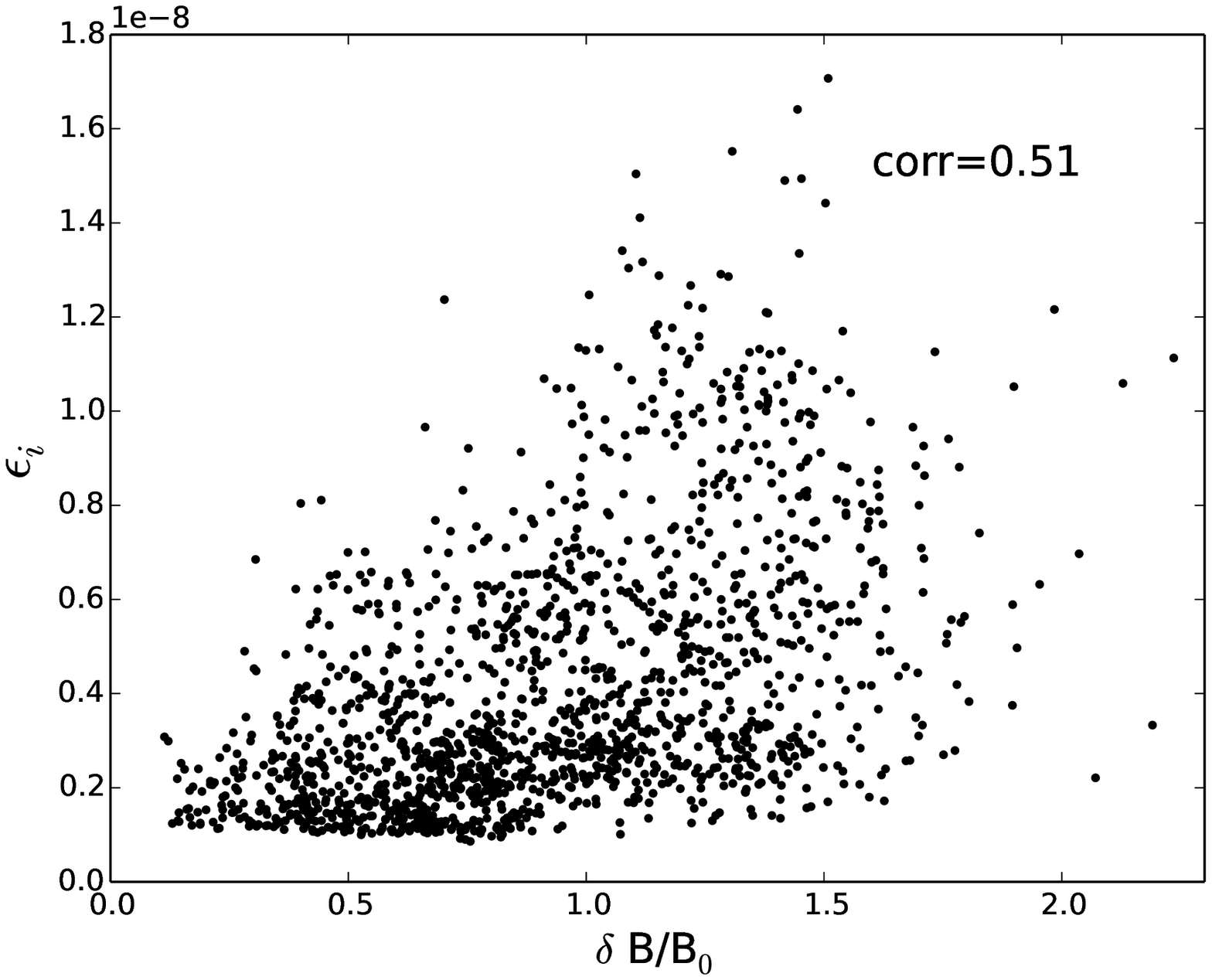}
\includegraphics[width=15pc]{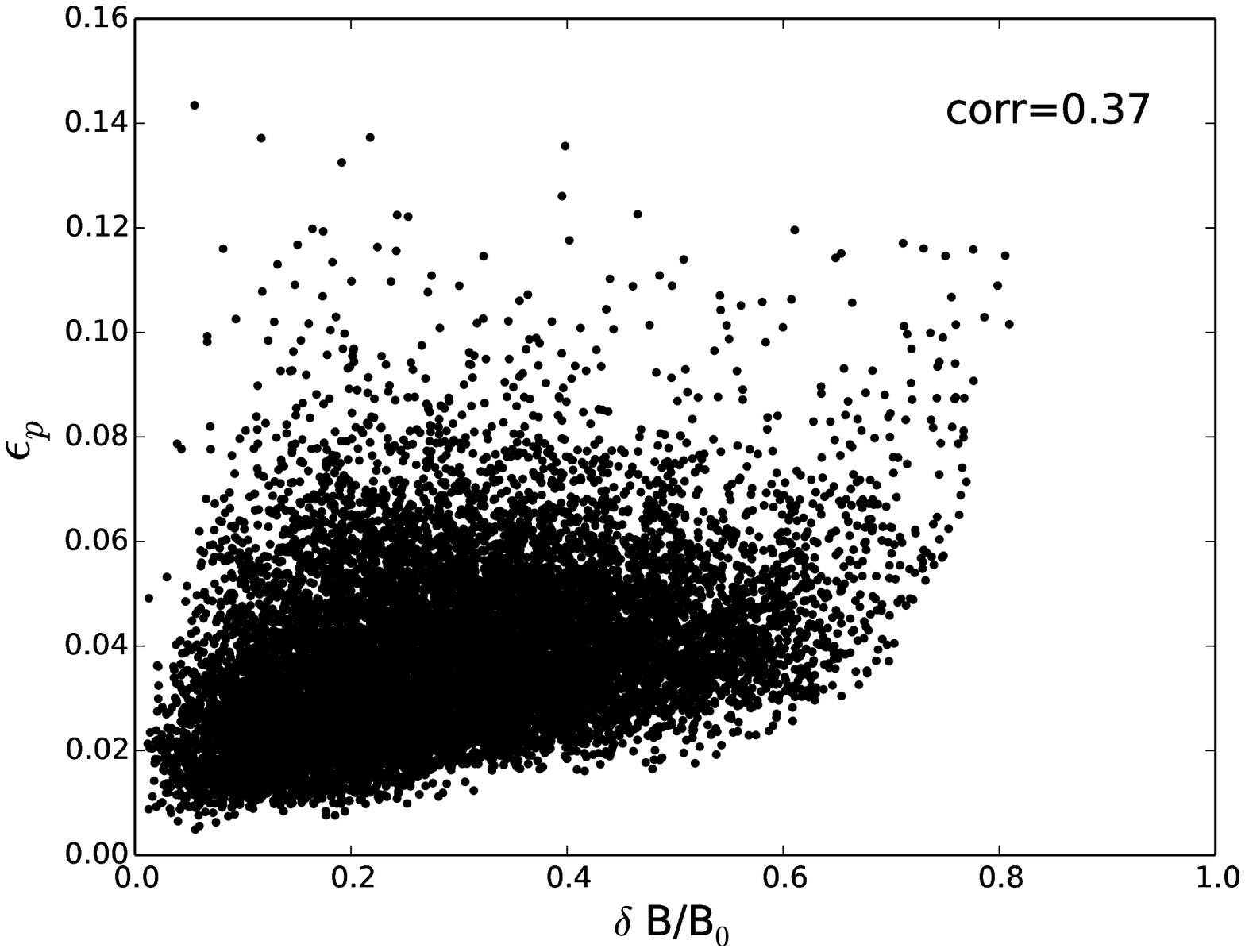}
  \caption{Same format as figure~\ref{fig:eps-fields} for the $\epsilon_i$ parameter as a function of the magnetic field fluctuations. Notice the good correlation both in the MMS data and in the numerical experiment.}
\label{fig:eps-dBB0}
\end{figure*}

\begin{figure*}
\centering
\includegraphics[width=15pc]{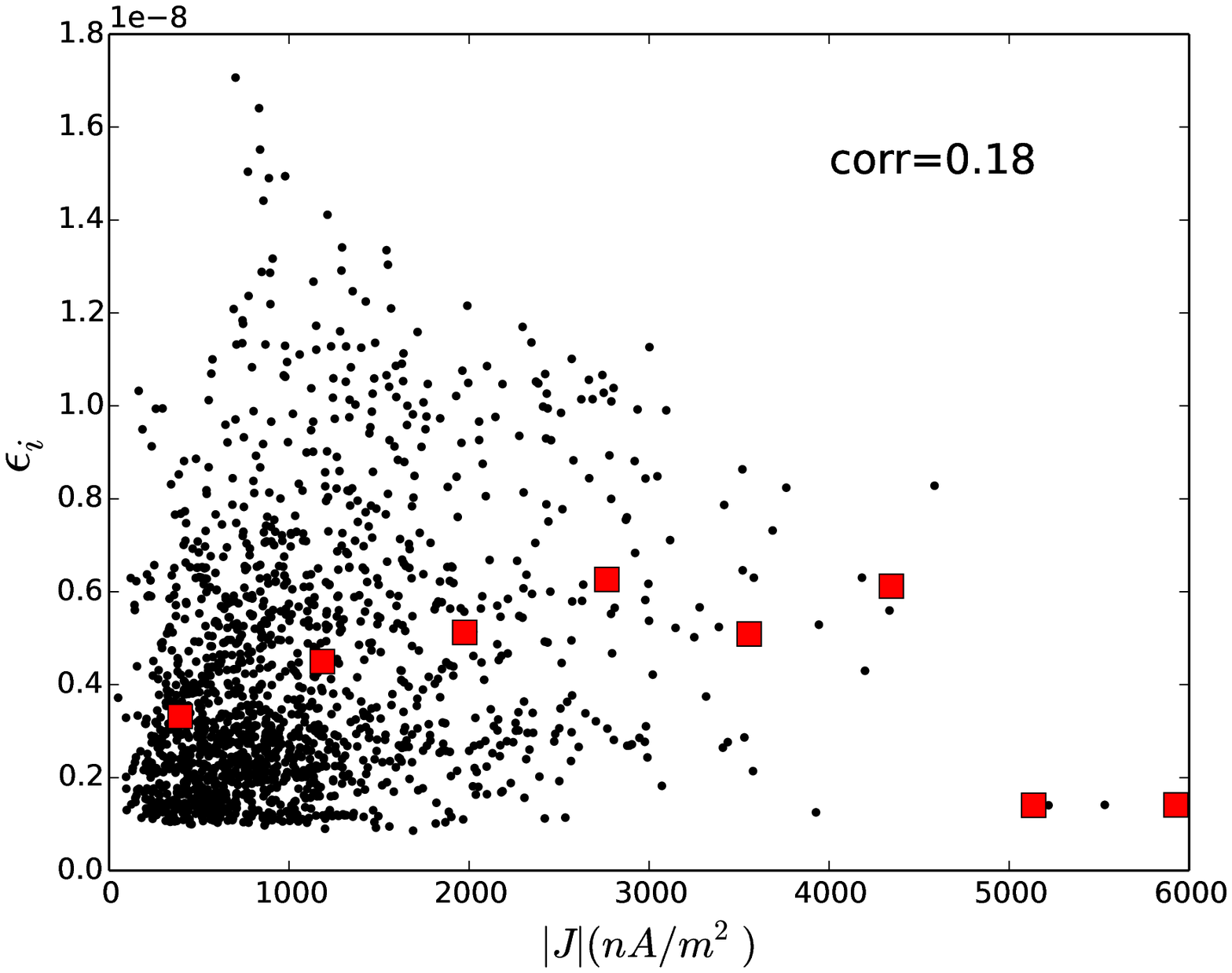}
\includegraphics[width=15pc]{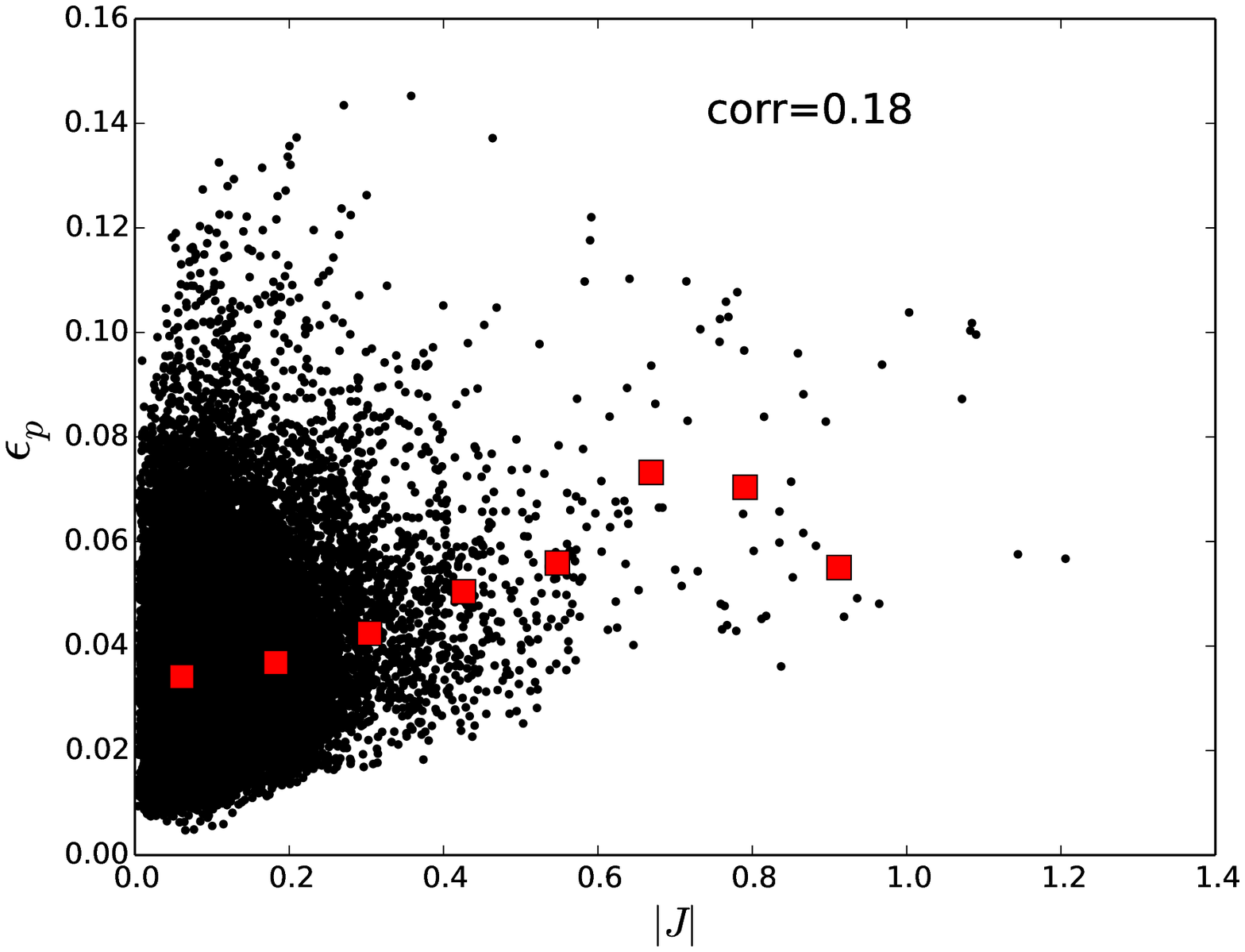}
 \caption{Same as figure~\ref{fig:eps-fields} for the $\epsilon_i$ parameter as a function of the current density magnitude.}
\label{fig:eps-current}
\end{figure*}

\begin{figure*}
\centering
\includegraphics[width=15pc]{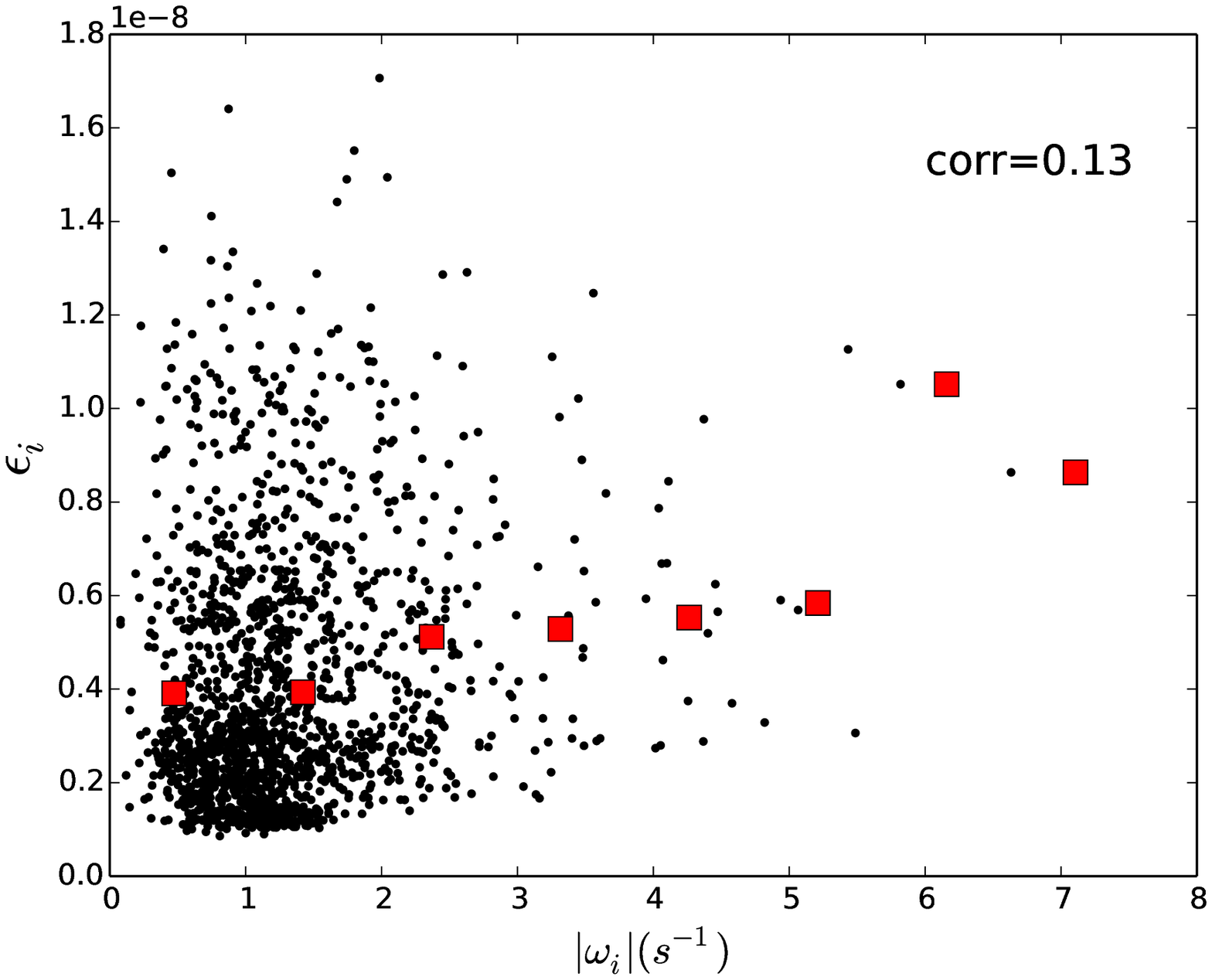}
\includegraphics[width=15pc]{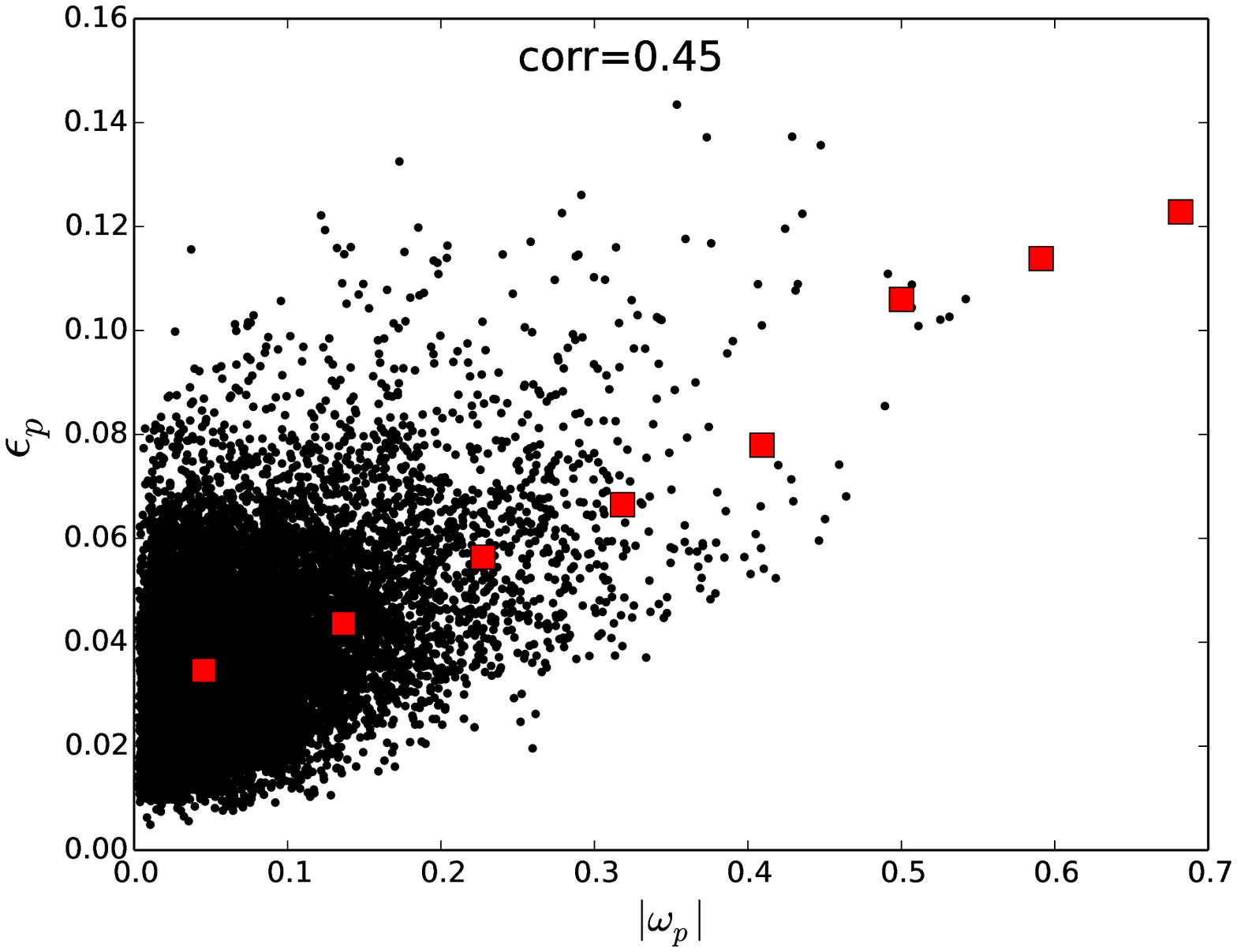}
 \caption{Same format as figure~\ref{fig:eps-fields} for the $\epsilon_i$ parameter as a function of the magnitude of the ion vorticity.}
\label{fig:eps-omega}
\end{figure*}

\begin{figure*}
\centering
\includegraphics[width=15pc]{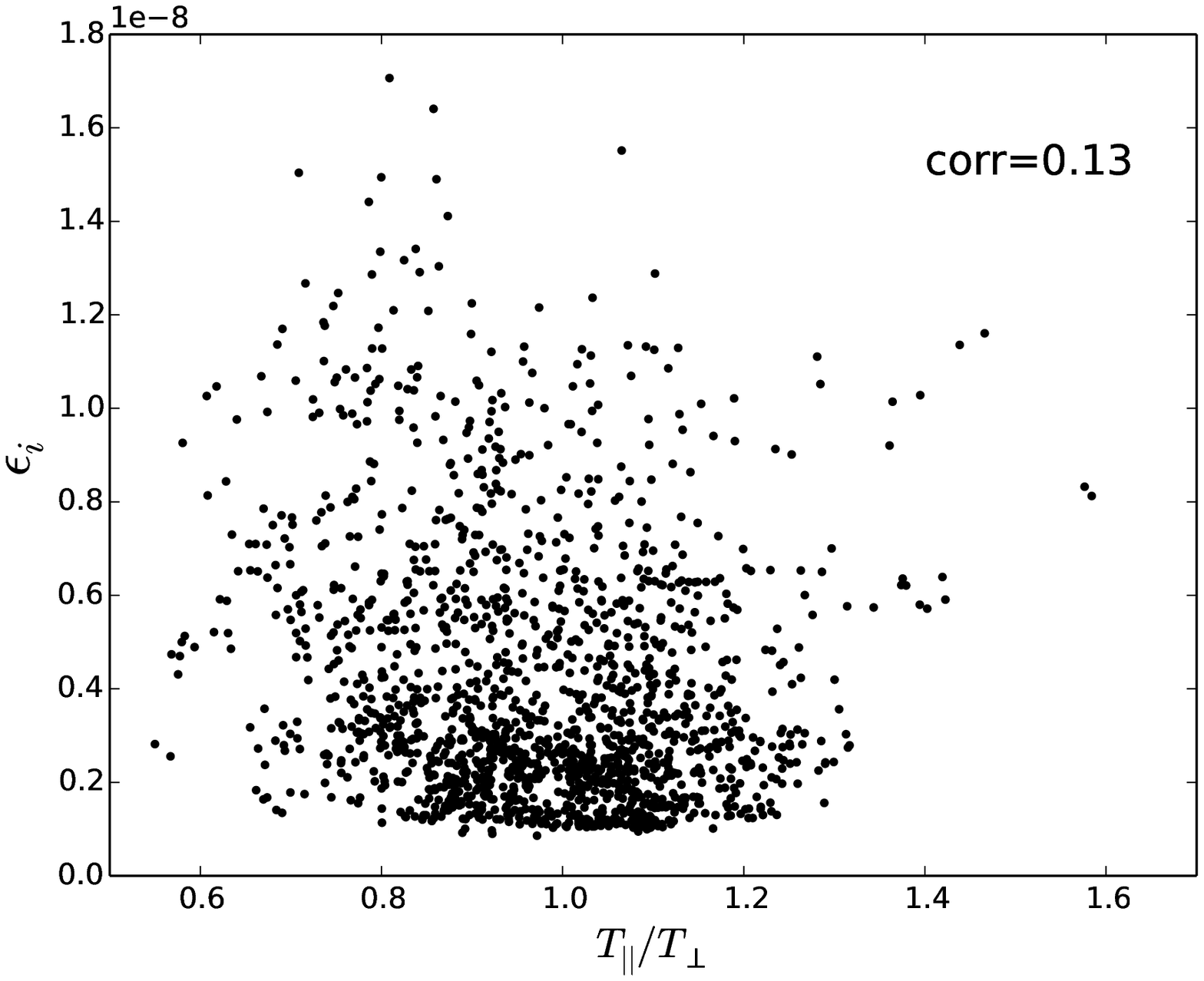}
\includegraphics[width=15pc]{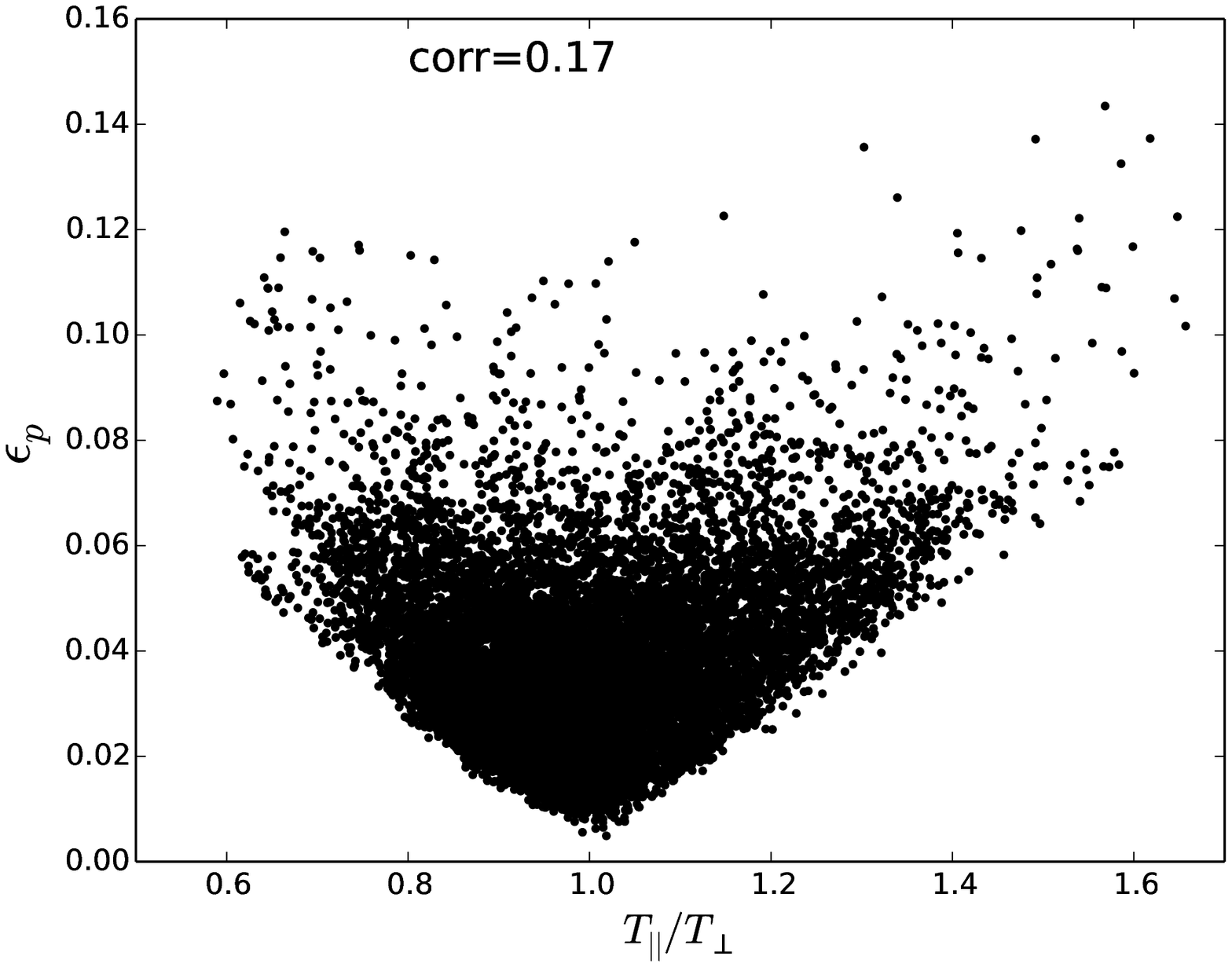}
 \caption{Same as figure~\ref{fig:eps-fields} for the $\epsilon_i$ parameter as a function of the temperature anisotropy.}
\label{fig:eps-tempaniso}
\end{figure*}

As noticed in figure \ref{fig:eps}, the time series of the $\epsilon_i$ parameter computed with the MMS1 data has some features that can also be recognized in the time series of other quantities, such as the perpendicular component of the electric field and the ion vorticity. In order to highlight such similarities, we have generated scatter plots of $\epsilon_i$ as a function of different quantities, in order to investigate the physical quantities that might correlate with distortions in the ion VDFs (both in the data and in the simulations). Figure \ref{fig:eps-fields} shows the scatter plots of $\epsilon_i$ versus the intensity of the electric field in the plasma frame, namely $\mathbf{E}'=\mathbf{E}+(\mathbf{V}_i\times \mathbf{B})$ degraded to $0.15$ sec resolution in the magnetosheath interval (left panel) and in the HVM simulation (right panel). The Pearson correlation coefficient is also reported in the panels, indicating a very good degree of correlation in the MMS1 data set. 
Of course, since the region is highly turbulent both the fields and the plasma quantities display considerably fluctuations, giving rise to a certain degree of dispersion in the scatter plots. Thus, we have overplotted the $\epsilon_i$ parameter mean values within bins of the electric field (red squares in figure \ref{fig:eps-fields}). This procedure highlights the correlation between the two quantities.
Besides the good degree of correlation with the intensity of the electric field, the $\epsilon_i$ parameter tends to be linearly correlated with magnetic field fluctuations $\delta B/B_0$ (see figure \ref{fig:eps-dBB0}) both in the data and in the simulation. This confirms that high distortions in the ion VDFs can occur within regions of strong magnetic activity \citep{Servidio12}.
On the other hand, figure \ref{fig:eps-current} displays a scatter plot of $\epsilon_i$ with the magnitude of the current density (see Section \ref{sec_event}) for the magnetosheath data (left panel) and for the HVM simulation (right panel). The Pearson correlation is $< 0.2$ in both cases and very weak correlation can be noticed from observing the the average $\epsilon_i$ values (red squares). Notice that \citet{Valentini16} have pointed out that in the simulations the deviation from the Maxwellian equilibrium is non-homogeneous in space and tends to be maximized around the peaks of the current density that naturally form at the interfaces of magnetic flux tubes. Thus, the good match between data and simulation in figure \ref{fig:eps-current} suggests that a similar scenario can be envisaged in the turbulent magnetosheath interval.
Besides the current density, the correlation between the ion vorticity, as computed with a multi-spacecraft technique in the MMS data (see Section \ref{sec_event}), and $\epsilon_i$ has been studied both in the data and in the simulation. The scatter plots are given in figure \ref{fig:eps-omega}. In such case, the correlation in the data is poor ($0.13$) although the average of $\epsilon_i$ in bins of $|\omega_i|$ shows a certain degree of correlation, while it is higher in the HVM simulations where the formation of vortices tends to  distort the ion VDFs. However, a closer look at figure \ref{fig:eps} points out that some peaks in $\epsilon_i$ occur during peaks in the ion vorticity, although the great variability of all those quantities at very short time scales makes the correlation analysis complex.
$\epsilon_i$ does not seem to correlate neither with ion temperature (not shown) nor with the temperature anisotropy (see figure \ref{fig:eps-tempaniso}) both in the data and in the simulation, as well as with the magnetic energy conversion/dissipation computed via $\mathbf{E}'\cdot \mathbf{J}$ (not shown). 
The fact that there is no evident correlation between $\epsilon_i$ and the temperature anisotropy, which intrinsically is a deviation from the thermodynamic equilibrium, suggests the emergence of a very complex scenario in the process of the turbulent energy transfer at ion and sub-ion scales related to phase-space structures that cannot be described in terms of pressure tensor anisotropy~\citep{CGL56}. 

\begin{figure*}
\centering
\includegraphics[width=15pc]{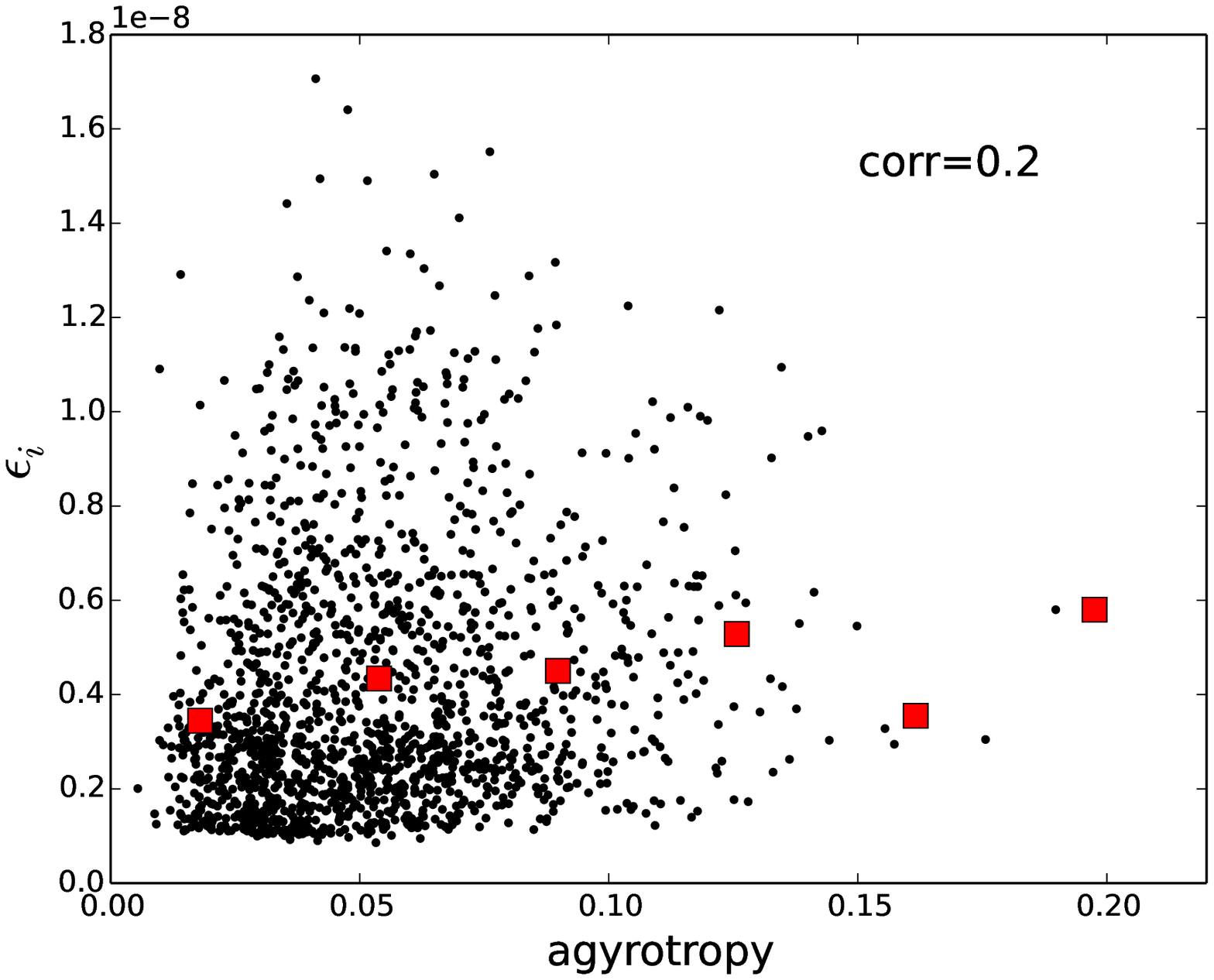}
\includegraphics[width=15pc]{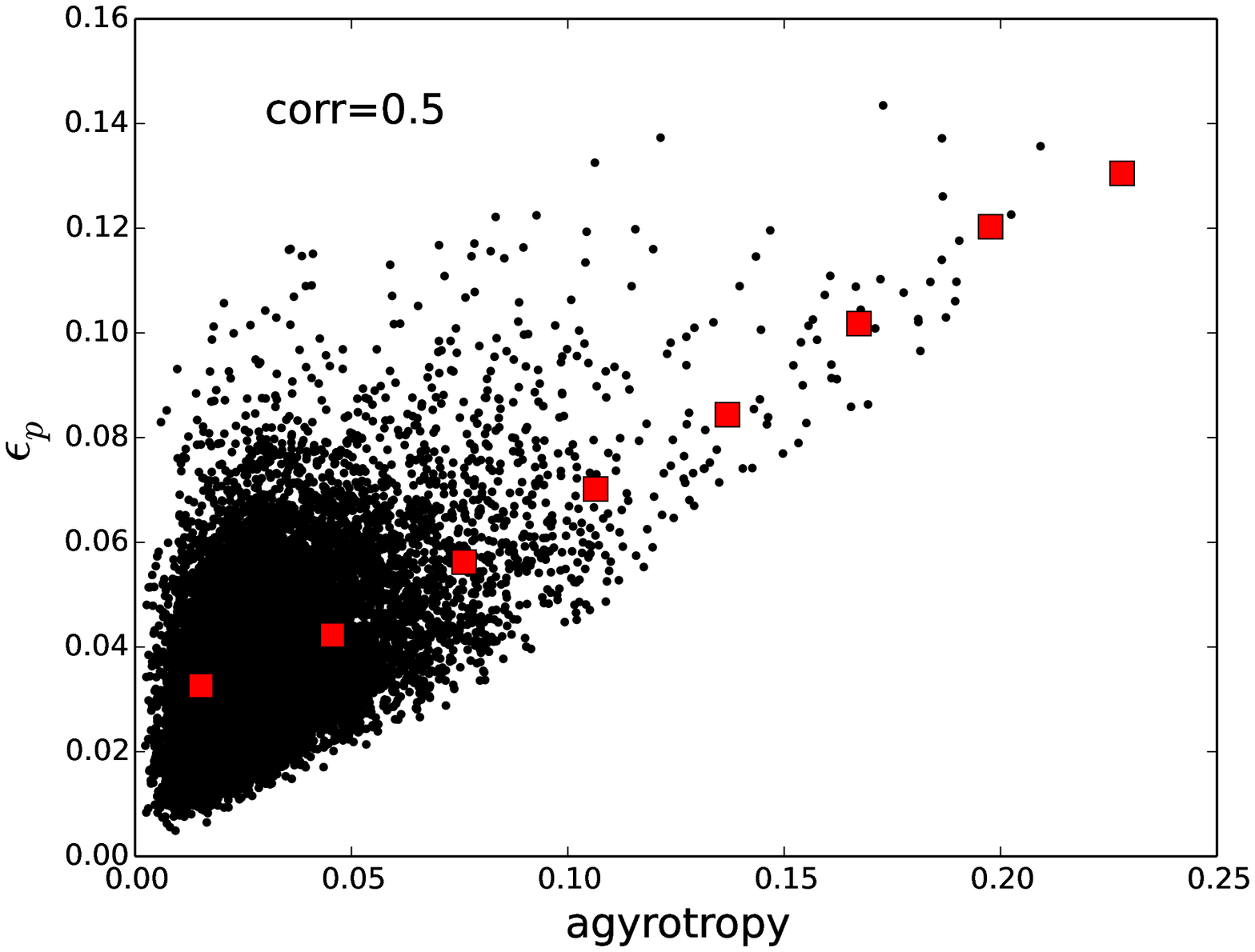}
 \caption{Same format as figure~\ref{fig:eps-fields} for the $\epsilon_i$ parameter as a function of the agyrotropy of the pressure tensor (see text).}
\label{fig:eps-agyrotropy}
\end{figure*}

The turbulent activity produces distortions of the VDFs that are much more complex than the generation of temperature anisotropy, thus hiding the natural correlation between $\epsilon_i$ and $T_\parallel/T_\perp$.
Here, the ion agyrotropy of the pressure tensor, namely the departure of the pressure tensor from cylindrical symmetry about the local mean field, has been calculated by using the expression of \citet{Swisdak16}
\begin{equation}
Q=1-\frac{4l_2}{(l_1-P_{||})(l_1+3P_{||})},
\label{agyrotropy}
\end{equation}
where $P_{||}=b^2_xP_{xx}+b^2_yP_{yy}+b^2_zP_{zz}+2(b_xb_yP_{xy}+b_xb_zP_{xz}+b_yb_zP_{yz})$, $l_1$ is the trace of the pressure tensor and $l_2=P_{xx}P_{yy}+P_{xx}P_{zz}+P_{yy}P_{zz}-(P_{xy}P_{yx}+P_{xz}P_{zx}+P_{yz}P_{zy})$. The correlation with $\epsilon_i$ is plotted in figure \ref{fig:eps-agyrotropy} and by definition $Q=0$ indicates gyrotropy. In the HVM simulation, the agyrotropy of the ion pressure tensor is well correlated with the deviation from Maxwellian. Thus, highly distorted ion VDFs have high probability to be associated with large agyrotropy. This is not really surprising since the latter can be seen as a deviation from Maxwellian. On the other hand, the correlation in the MMS1 data (left panel in figure \ref{fig:eps-agyrotropy}) is much weaker than the one observed in the simulations, although again a certain linear trend can be recognized. 

\begin{figure*}
\centering
\includegraphics[width=15pc]{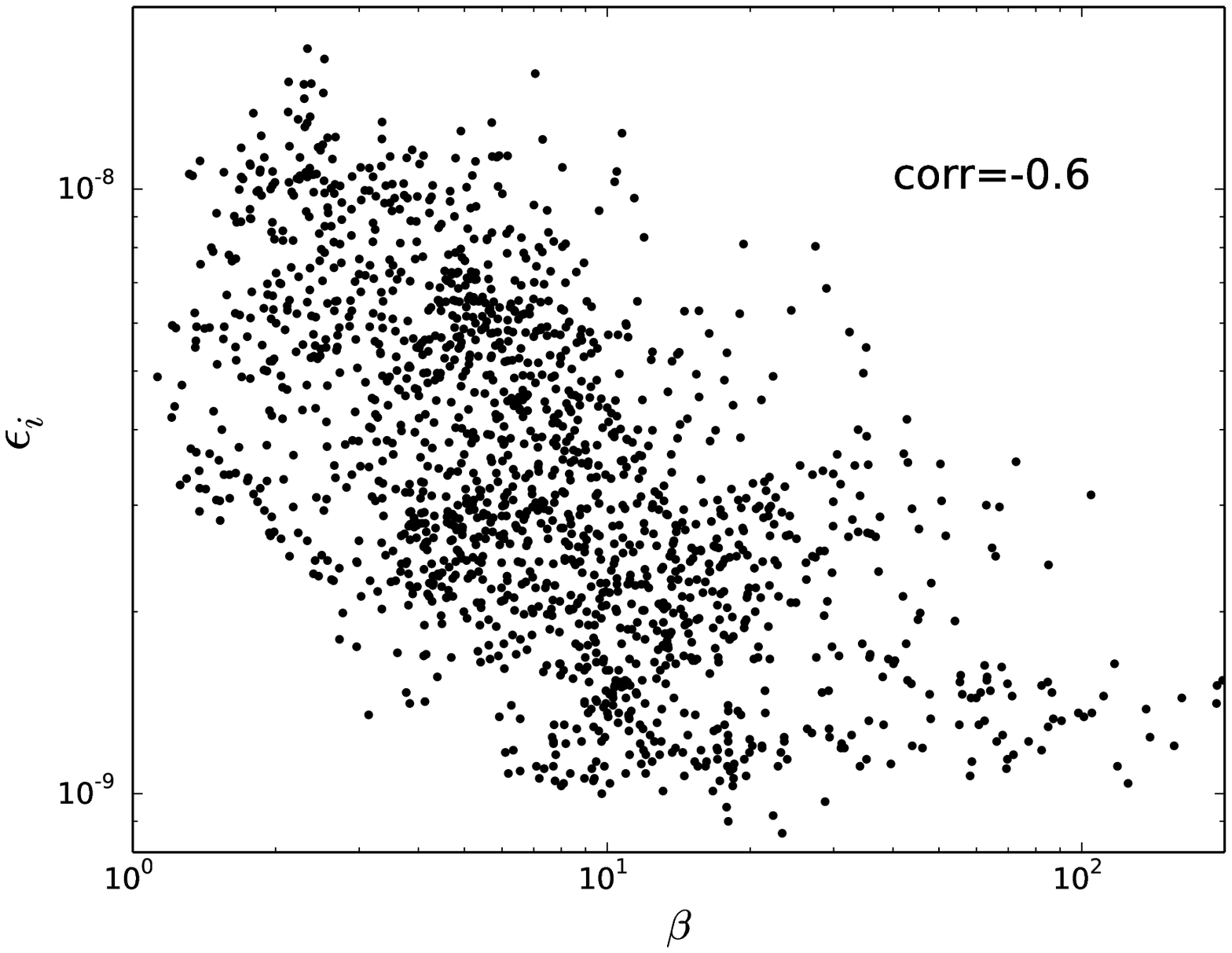}
 \caption{Scatter plot in log-log axes of the $\epsilon_i$ parameter as a function of the plasma beta in the MMS1 data set. A clear anticorrelation (computed using the Spearman correlation) has been found.}
\label{fig:eps-beta}
\end{figure*}

Finally, the degree of correlation between $\epsilon_i$ and the plasma $\beta$ in the MMS1 data set has been studied. The plasma beta varies over a broad range of values and tends to be greater than $1$. Thus, we have reported in figure \ref{fig:eps-beta} the scatter plot in log-log axis between $\epsilon_i$ and $\beta$: a good anticorrelation can be recognized (with a Spearman correlation coefficient $\sim -0.6$). In this case, since the range of variation of $\beta$ is pretty large, the linear Pearson correlation is much lower than the Spearman. The largest deviations from Maxwellian clusterize around $\beta<3$, that is when the thermal speed of ions tends to be of the same order of the Alfv\'en speed. In such a case, all the fluctuations that propagate at $v_A$ interact with the bulk of the ion VDF, inducing distortion in the distribution function.


\section{Discussions}\label{conclusions}
The highly turbulent medium downstream of the quasi-parallel bow shock, as observed by the MMS spacecraft, shows a very bursty behavior in both magnetic field and plasma data. The parameter $\epsilon_i$, which quantifies the degree of deviation of the ion VDFs from a Maxwellian shape, has been computed throughout the 5 minute interval analyzed, showing again high variability, with regions of strong departure of VDFs from thermodynamic equilibrium. Such a great variability makes the analysis of the correlation between $\epsilon_i$ and physical quantities, such as temperature anisotropy, current density, ion vorticity, very complex.
A comparison between MMS1 data and results coming from a HVM simulation has been performed. The comparison highlights a certain degree of correlation between peaks in the $\epsilon_i$ (high deviation from a Maxwellian shape) and the presence of high current density (thin current sheets) regions. However, no clear correlation has been found with temperature anisotropy (calculated with respect to the local magnetic field) and with the magnetic energy conversion/dissipation (taken as $\mathbf{E}'\cdot \mathbf{J}$) both in the data and in the simulations. Similar results (not shown) have also been found in a MMS data set on 8 September 2015, when the spacecraft was located in the dusk-side magnetopause, moving towards the magnetosheath \citep{Stawarz16}.

The correlation between $\epsilon_i$ and the ion temperature is rather complex. Indeed, a more in-depth analysis gives indication of a temperature increase in the vicinity of large amplitude peaks in epsilon (selected, for example, via a threshold method), but not at the peaks. This would justify a more pronounced anticorrelation observed in the data (not shown).
Numerical simulations also exhibit no correlation between distortion of the proton VDF and proton temperature. Recently, \citet{Chasapis18} have shown a strong parallel electron heating at regions of high current density also associated to distortion of the electron VDFs.
The analysis carried out on the ions shows that the presence of localized regions of high current density and strong energy dissipation (by means of $\mathbf{E}'\cdot \mathbf{J}$) weakly influences the shape of the particle distribution functions, as also displayed in numerical experiments \citep{Valentini16,Sorriso2018}, thus making complex and challenging the determination of the origin of non-Maxwellianity of ion VDFs in collisionless plasmas~\citep{Sorriso2019}. 
On the other hand, there is a clear tendency to develop non gyrotropic features close to velocity gradients in the plasma~\citep{Franci2016,Parashar2016,Valentini16,Sorriso2018}, probably due to a shear-induced anisotropization mechanism~\citep{DelSarto2016,DelSarto2018}, within regions with $\beta\sim 1$, where the interaction between the bulk of the ion VDF and the fluctuations propagating at $v_A$ is maximized, and close to strong bursts of perpendicular electric field, being ions highly magnetized~\citep{Sorriso2019}. 
It is worth stressing that at sub-ion scales the energy stored in the electric fluctuations is higher than that contained in the magnetic field fluctuations. This evidence, together with the observation of a correlation between $\epsilon_i$ and the electric field in the plasma frame, suggests that the nature of the distortion of ion VDFs can be electrostatic. However, this aspect needs further investigation and opens to the possibility to a strong interaction between particles and almost electrostatic waves, propagating along the mean magnetic field, at sub-ion scales, as a mechanism for the energy cascade towards the electron scales. 


\acknowledgements
This work has received funding from the European Unions Horizon 2020 research and innovation programme under grant agreement No 776262 (AIDA, www.aida-space.eu). EY was supported by the Swedish Civil Contingencies Agency, grant 2016-2102. DP was supported by STFC grant ST/N000692/1.



\bibliographystyle{jpp}

\bibliography{non-maxwellianity_3.bbl}

\end{document}